\documentclass[usenatbib]{mn2e}

\usepackage{graphicx}
\usepackage[usenames]{color}
\usepackage{amssymb}
\usepackage[fleqn]{amsmath}
\usepackage{mathtools}
\usepackage{multirow}
\usepackage{rotating}
\usepackage{amsfonts}

\usepackage{pifont}

\usepackage{url}
\usepackage{hyperref}

\hypersetup{
  colorlinks = true,
  pdftitle   = {Feedback Concordance in SPH simulations},
  pdfauthor  = {F. Durier, C. Dalla Vecchia},
  linkcolor  = magenta,
  citecolor  = blue,
  filecolor  = green,
  urlcolor   = blue
}

\newcommand{\gadgettwo}{\textsc{gadget}-2}
\newcommand{\gadget}{\textsc{gadget}~}

\newlength{\colwidth}
\begin{document}

\title[Towards concordance of feedback methods in SPH]{Implementation of feedback in SPH: towards concordance of methods}

\author[F. Durier \& C. Dalla Vecchia] {Fabrice Durier$^1$\thanks{E-mail: fdurier@mpe.mpg.de} and Claudio Dalla Vecchia$^1$\thanks{E-mail: caius@mpe.mpg.de}\\ 
$^1$Max Planck Institute for Extraterrestrial Physics, Gissenbachstra\ss{}e 1, 85748 Garching, Germany}

\maketitle

\abstract 
We perform simulations of feedback from supernovae and black holes with smoothed particle hydrodynamics (SPH). 
Such strong perturbations are inaccurately handled with standard time integration schemes, leading to poor energy conservation, a problem that is commonly overlooked.
We show for the first time that, in the absence of radiative cooling, concordance of thermal and kinetic feedback are achieved when using an accurate time integration. 
In order to preserve the concordance of feedback methods when using a more efficient time integration scheme -- as for instance the hierarchical time-step scheme -- 
we implement a modified version of the time-step limiter proposed by \cite{Saitoh2009}.
We apply the limiter to general test cases, and first show that this scheme violates energy conservation up to 
almost four orders of magnitude when energy is injected at random times.
To tackle this issue, we find necessary, not only to ensure a fast information propagation, 
but also to enforce a prompt response of the system to the energy perturbation.
The method proposed here to handle strong feedback events enables us to achieve energy conservation at percent level in all tests,  
even if all the available energy is injected into only one particle.
We argue that concordance of feedback methods can be achieved in numerical simulations only if the time integration scheme preserve a high energy conservation level.
\endabstract


\section{Introduction}

In the last decades, it has become clear that energetic feedback processes are the key ingredients in shaping 
the star formation history of the Universe and regulating the formation and evolution of galaxies. 
In particular, supernovae (SNe) and black holes (BHs) are considered the principal sources of such feedback \citep[see][for a review]{Silk2007}.

Cosmological simulations have been one of the most powerful tool for investigating the formation and evolution of galaxies \citep[see review from][]{Bertschinger1998,SpringelRev2010}.
However, current cosmological simulations lack both the resolution and the physics needed to model small scale feedback phenomena like the explosion of a single SN or the accretion into a BH.
Several numerical, sub-grid recipes have been employed to mimic the large scale effect of feedback by injecting the energy either in thermal or kinetic form.

Thermal feedback has the advantage of having an isotropic effect on the closest surrounding medium. 
However, in the presence of radiative cooling and poor resolution, a large part of the energy is radiated away before any hydrodynamical response of the medium. 
Several studies concentrated their efforts in numerically solving the over-cooling problem and the consequent dissipation of thermal feedback energy 
\citep[e.g.][]{Gerritsen1997,Mori1997,Thacker2000,Kay2002,Sommer-Larsen2003,Brook2004,Stinson2006,Booth2009}.
In order to avoid the over-cooling problem, other studies favoured the kinetic feedback approach, in which all or part of the energy is given to the surrounding medium as momentum 
\citep[e.g.][]{Navarro1993,Mihos1994,Kawata2001,Kay2002,Springel2003,Oppenheimer2006,DallaVecchia2008,Dubois2008}. 
However, over-cooling can still play an important role in quickly dissipating the heat produced by the outflowing gas shocks, and underestimating the increase of the temperature of the medium.

Until now, despite the large variety of implementation of feedback models, very little work has been done to compare 
the two feedback approaches in the context of galaxy formation.
\cite{Kay2002} compared the results of cosmological simulations with either thermal or kinetic SN feedback. 
They showed that even when artificially preventing cooling of heated particles the results of the two schemes are considerably different.
Nevertheless, before tackling the over-cooling problem, one should first make sure that the input energy is accurately conserved.

\begin{figure}
  \begin{center}
    \hspace{-2mm}\includegraphics[trim = 0mm 2mm 2mm 4mm, clip, width=0.24\textwidth]{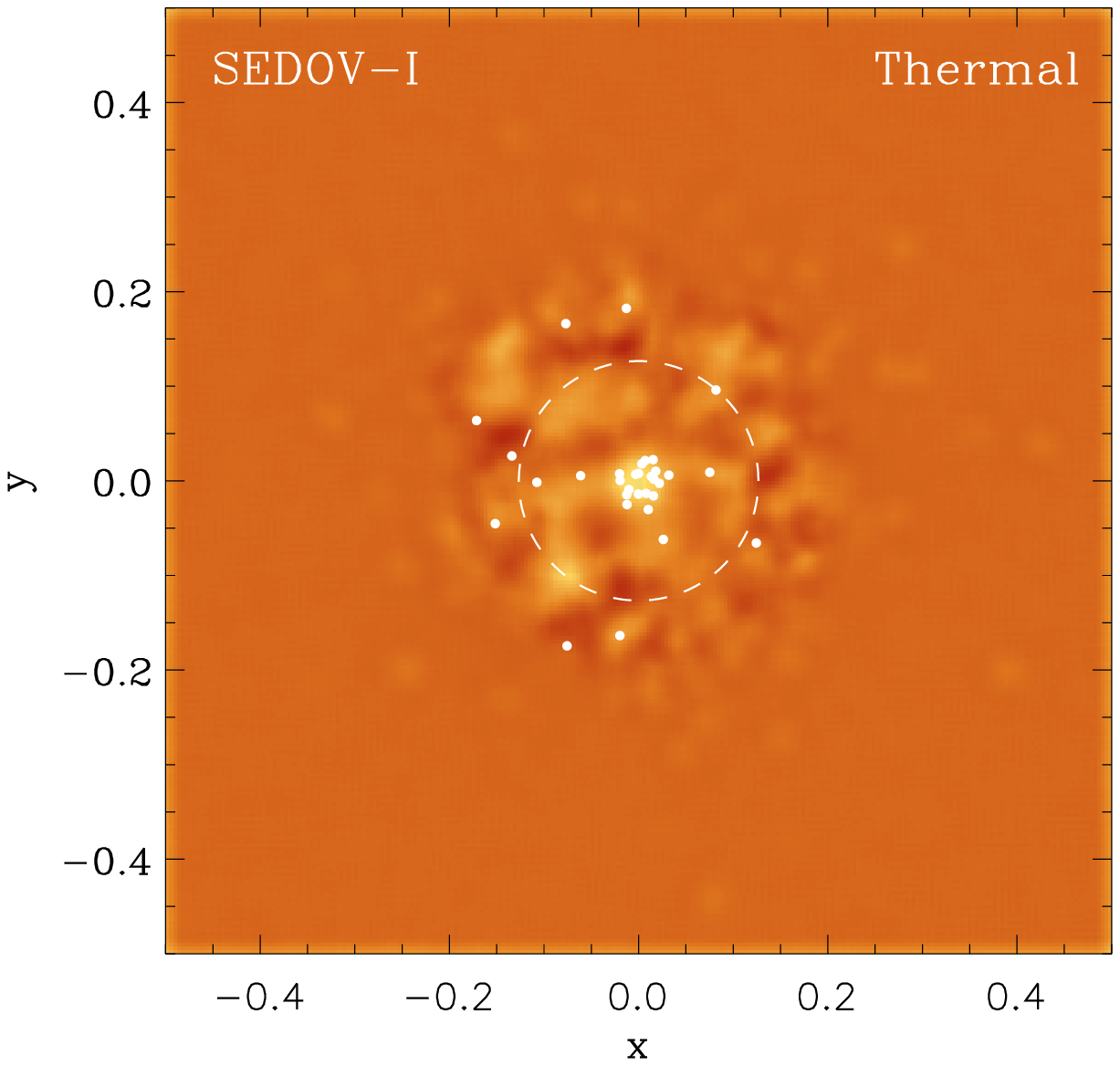}
    \hspace{-1mm}\includegraphics[trim = 0mm 2mm 2mm 4mm, clip, width=0.24\textwidth]{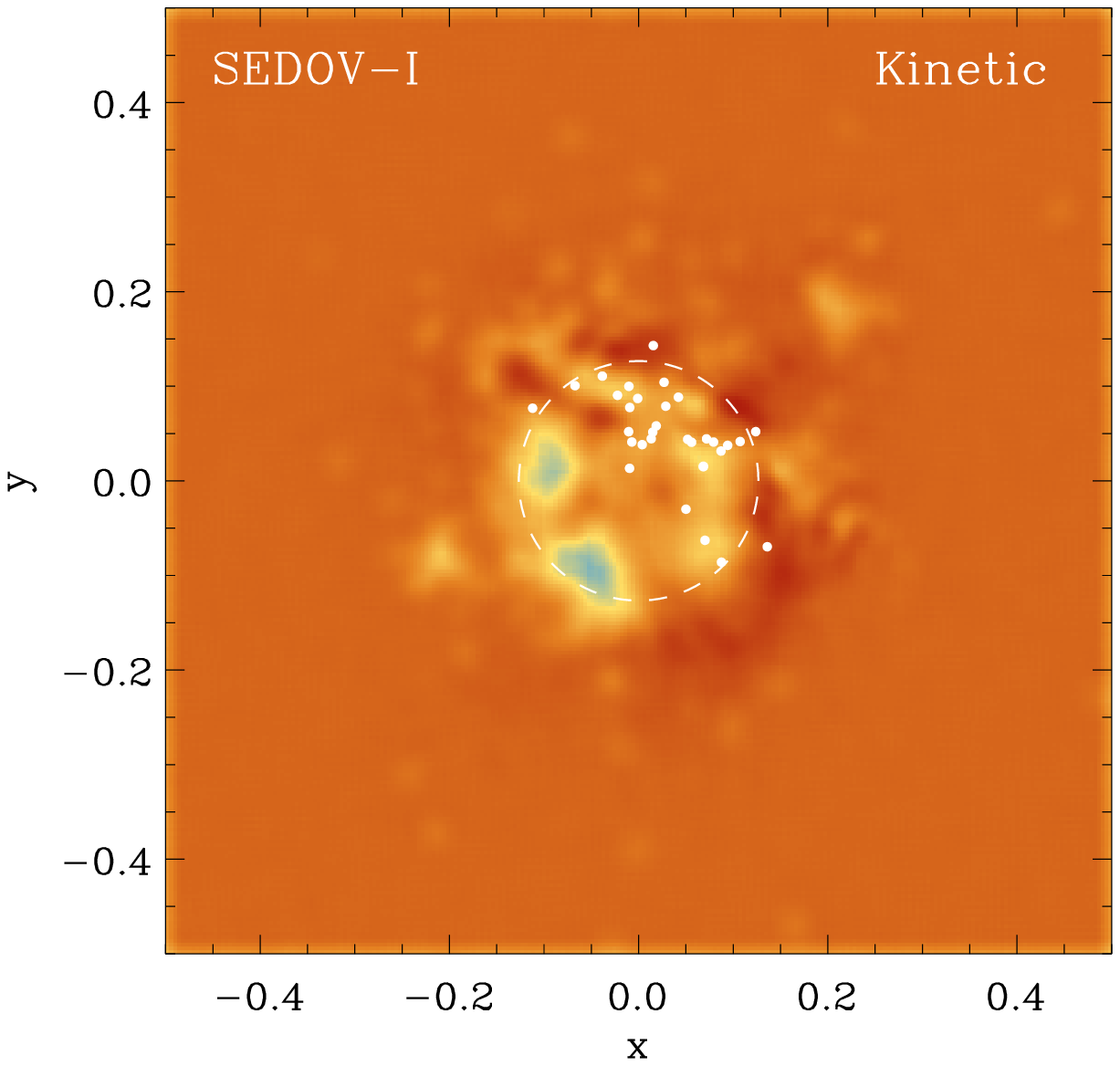}\\
    \hspace{-2mm}\includegraphics[trim = 0mm 4mm 2mm 4mm, clip, width=0.24\textwidth]{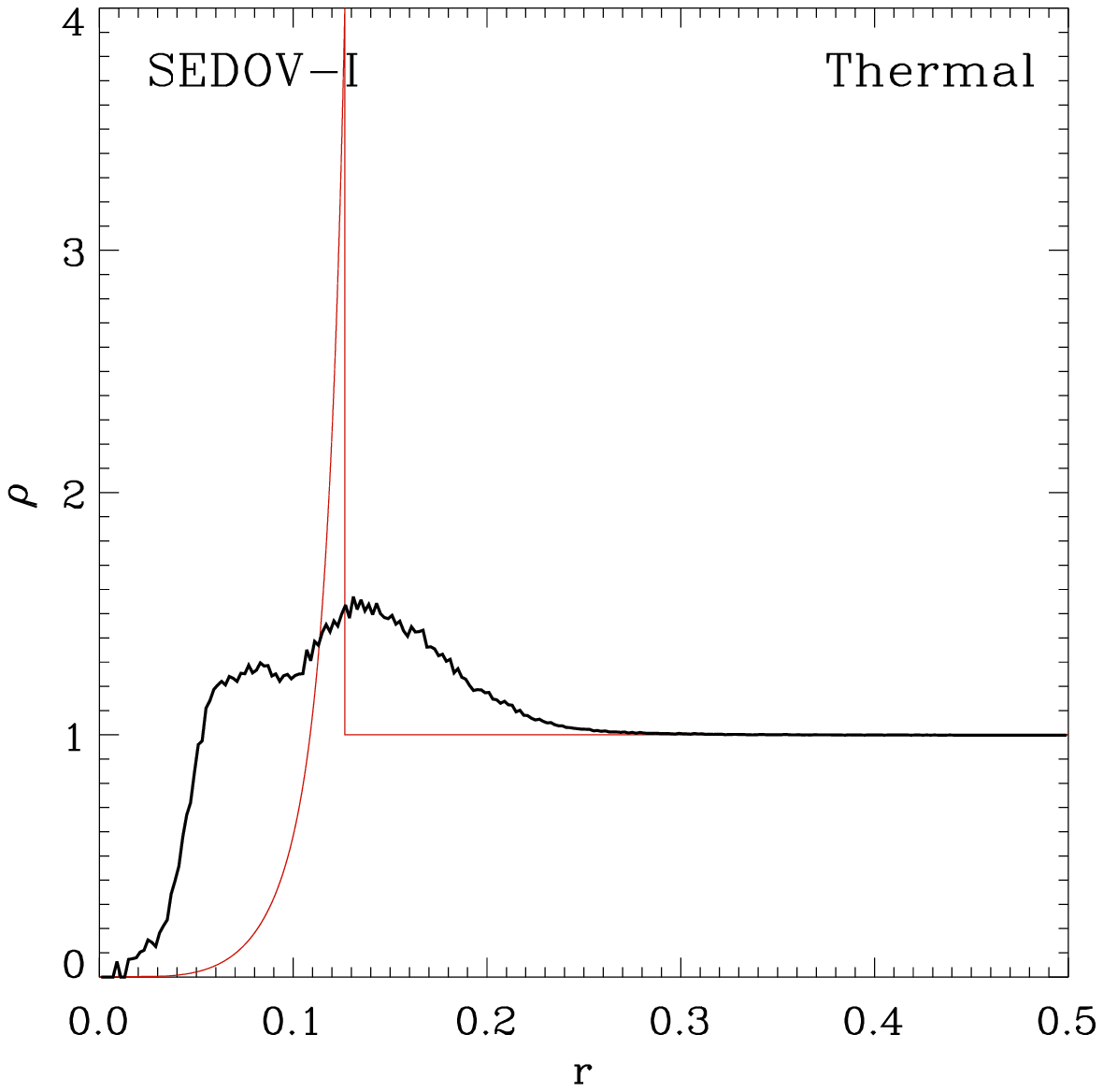}
    \hspace{-1mm}\includegraphics[trim = 0mm 4mm 2mm 4mm, clip, width=0.24\textwidth]{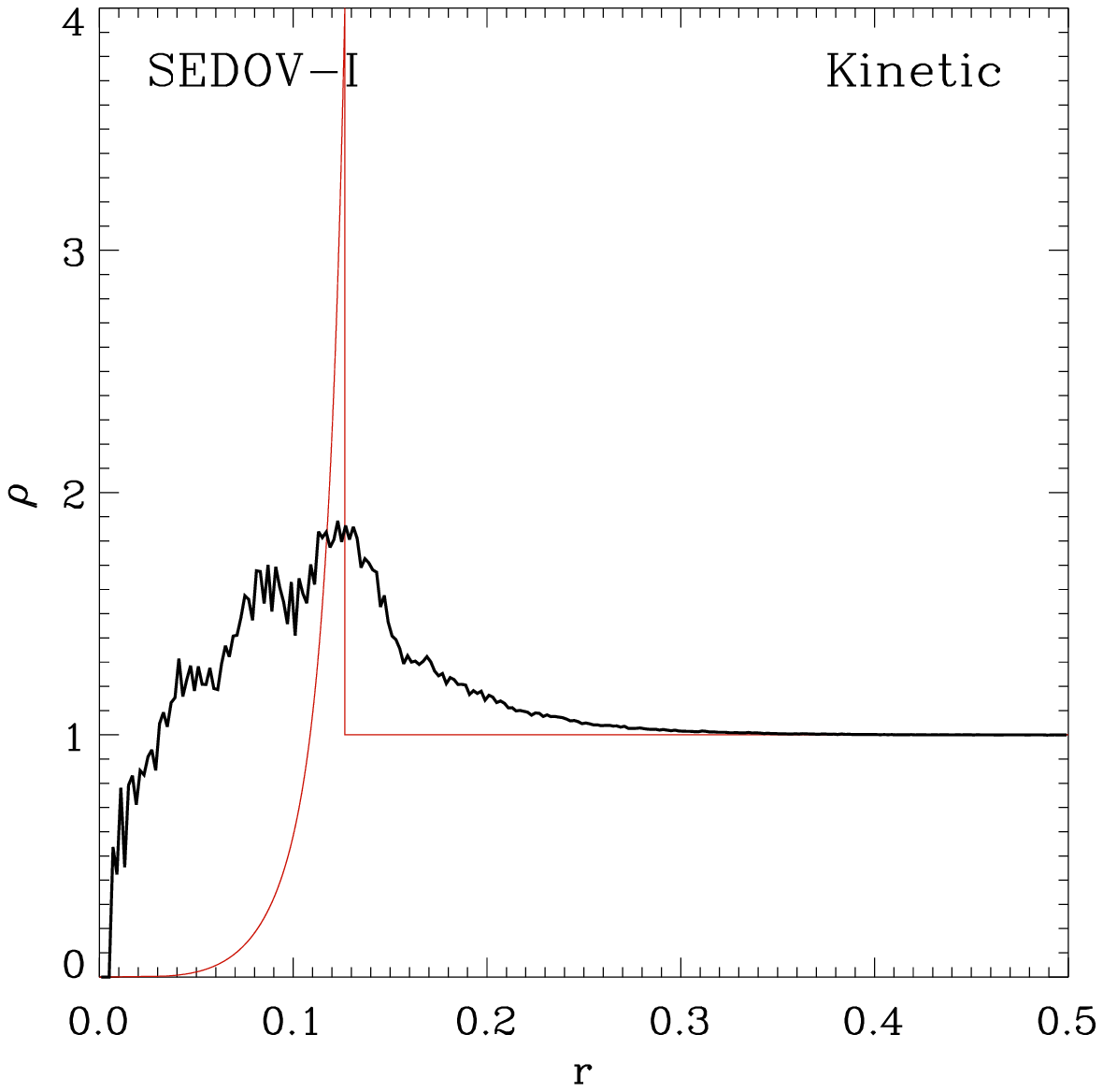}
  \end{center}
  \caption{Sedov's explosion test for thermal (left column) and kinetic (right column) energy injection, 
	when the individual time-step integration is used. Results are given in the natural system of units soon after the explosion at time $t=0.004$.
	We show in the upper panels the projected gas density in a slice of thickness $\Delta z=0.1$ centred on the box origin. 
	The dashed circle corresponds to the position of the spherical shock-front given by the similarity solution in Eq.~\ref{eq:sedovblast} at that time.
	The white dots mark the position of all particles that initially received the energy.
	In the lower panels, the average radial density profiles are compared to the profiles given by Sedov's analytic solution (red lines).
	Given the large energy jump introduced by the explosion, it is clear from these plots that none of the feedback scheme are able to describe properly Sedov's test.
  \label{fig:sedovindiv}}
\end{figure}

In a recent study of the modelling of SN feedback in SPH simulations, \cite{Saitoh2009} \citep[hereafter SM09, see
also][]{Merlin2010} noted that strong perturbations of the internal
energy of gas particles lead to the violation of energy conservation
when an individual particle time-step scheme is used (see Fig.~\ref{fig:sedovindiv} for an illustration).
That can be summarised as follows. If a source of
energy alters the internal energy of a (active) gas particle, the
particle will eventually decrease its time-step according to the Courant criterion. However, if its
neighbouring particles are inactive,\footnote{Their hydrodynamical
state is not updated at the same time.} they will not react
immediately to the change of the thermal state of the
region, and the integration accuracy may be strongly compromised. The lack of a
prompt response to the nearby energy injection can lead to effects
like inter-particle crossing (due to particles missing the deceleration given by
viscous forces) and non-conservation of energy.

SM09 proved that, when using an appropriate time-step limiter, accurate energy
conservation is achieved for strong internal energy perturbations.
They proposed a scheme in which the ratio of the time-steps (longer over shorter) of
neighbouring gas particles cannot be larger than a given factor
$f_{\rm step}$. A value of $f_{\rm step}=4$ is enough to ensure
energy conservation to similar level of accuracy given by a global
time-step integration scheme (\cite{Merlin2010} suggested the values $f_{\rm
 step}=[4,8]$).

In this work and for the first time, we investigate under what conditions thermal and kinetic feedback schemes lead to qualitatively identical results. 
We first show that the feedback prescriptions are equivalent if the integration of the hydrodynamical equations is done over global, system time-steps. 
In order to do that, we compare test simulations of Sedov's blast wave problem \citep{Sedov1959,Landau1959} to its analytic solution, finding very good agreement and energy conservation to within 1\%.

With the aim of generalising the results, we investigate whether one can achieve similar agreement and accuracy 
with the more computationally efficient individual time-step integration scheme.
We implement a modified version of the SM09 time-step
limiter in the public version of \gadgettwo{} \citep{Springel2005} consistently taking into account the time-step synchronisation and the underlying leapfrog integrator.
We test the robustness and speed of the limiter with Sedov's explosion
problem injecting either thermal or kinetic energy at a random time.
We demonstrate that thermal and kinetic feedback methods are equivalent
and energy is accurately conserved, provided that
the system promptly responds to the energy perturbation.

With the application of the time-step scheme to galaxy formation and cosmological simulations in mind, we study a more realistic problem: the release of energy
in the presence of density and pressure gradients. We apply the scheme to an off-centre
explosion in a self-gravitating gas halo, and show that the
concordance of feedback methods and accurate energy conservation are still achieved even for the
extreme case where all available energy (either thermal or kinetic) is injected into only one particle.

In standard cosmological simulation practice the new integration approach presented in this paper 
could, in principle, require more simulation steps, but will mainly populate the intermediate time-bin levels with more particles. 
Therefore, we could expect a slight increase of the computational time for galaxy formation simulations,
a price to pay for a better handling of feedback events.
However, we still want to emphasise that, in idealised simulations like the ones presented in this work,
the proposed scheme enables much faster integration than the standard, individual time-step scheme 
since it preserve a high energy conservation accuracy.

The paper is organised as follows. We first show in Sec.~\ref{sec:concordance} that concordance
of feedback methods is achieved with a global time-step scheme.
We then apply our individual time-step algorithm to Sedov's blast wave
test and to the off-centre explosion in a self-gravitating gas halo in Sec.~\ref{sec:individual}.
We conclude in Sec.~\ref{sec:discuss}.

In all the simulations presented in this paper, we use the so-called \textit{natural} system of units in which the units of velocity, length and mass are unity.


\section{Concordance under accurate time integration}
\label{sec:concordance}

\bigskip

\begin{table*}
\begin{center}
\begin{tabular}{ l | l | c | c | cc | cc }
\hline
\hline 
\multirow{2}{*}{Name} & \multirow{2}{*}{Scheme} & \multirow{2}{*}{$E_{\star}$} & \multirow{2}{*}{$\alpha$} & 
\multicolumn{2}{c}{ Energy (\%) } & \multicolumn{2}{c}{ Time (h) } \\ 
& & & & Thermal & Kinetic & Thermal & Kinetic \\ 
\hline 
\hline 
{\bf SEDOV-G} & {\bf Global} & $\mathbf{1}$ & {\bf 2} & {\bf 0.81} & {\bf 0.78} & {\bf 3.71} & {\bf 2.50} \\ 
\hline 
SEDOV-G-$E01$ & Global & $0.1$ & 2 & 0.83 & 0.82 & 1.97 & 1.46 \\ 
SEDOV-G-$E001$ & Global & $0.01$ & 2 & 0.84 & 0.74 & 1.09 & 0.96 \\ 
\hline 
SEDOV-G-$\alpha\, 1$ & Global & $1$ & 1 & 1.99 & 2.30 & 3.87 & 2.73 \\ 
SEDOV-G-$\alpha\, 4$ & Global & $1$ & 4 & 0.73 & 0.61 & 3.58 & 2.38 \\ 
\hline 
\end{tabular}
\end{center}
\caption{Numerical parameters and energy conservation estimate and computational time for Sedov's blast wave tests with a global time-step scheme. 
From left to right: names of the runs specifying the integration scheme and the change of numerical parameters, time integration scheme; 
injected energy, $E_{\star}$; artificial viscosity, $\alpha$; energy conservation estimates in percent (from Eq.~\ref{eq:energyconservation}); and computational time in hours.
Statistics are given at the end of the runs, at time $t=0.1$.  The wall-clock time is given for runs on 8 cores. The reference simulation is highlighted in bold.}
\label{table:concordstat}
\end{table*}

In this section we demonstrate the concordance of thermal and kinetic feedback schemes. 
We run several simulation tests of Sedov's explosion problem varying the numerical parameters to show the robustness of our findings. 
Here, we assume the most favourable case in term of integration accuracy. 
The integration is done over global time-steps in order to concentrate only on how and why concordance is achieved.

We recall that Sedov's problem with thermal energy input has already been numerically solved with SPH \citep[e.g.][]{Springel2002,Tasker2008}, 
reaching the desired integration accuracy by limiting the particle maximum time-step or evolving the system on small, constant time-steps. 


\subsection{Simulation setup}
\label{subsec:globalsedovsetup}

\bigskip

We generate glass-like initial conditions \citep{White1996} with $N=128^3$ gas particles by evolving a random distribution of particles with inverted gravitational force sign.
We further relax the glass particle distribution by running it with hydro forces only, in order to smooth the pressure gradients that may remain due to the small density fluctuations. 
In the \textit{natural} system of units the size of the box is $L=1$. In order to obtain a uniform background density, $\rho_0=1$, the gas particle mass is set to $m_g=1/N$.
We assign to each particle the internal energy $u_0=10^{-3}$ so that the late evolution of the explosion is not affected by the energy of the medium 
that the expanding shell accumulates with time.

The total energy $E_{\star}$ is injected in a small region in the centre of the volume either in thermal or kinetic form.
In the simulations presented in this section, we heat/kick the $n_{\rm [h,k]}=32$ particles within a sphere of radius $r_0$ 
(we refer to heated and kicked particles with the subscripts `h' and `k', respectively). 
The energy is distributed using the SPH kernel weight normalised to unity, and with kernel size $r_0$ 
Particle $i$ at distance $r_i$ from the centre of injection receives the energy per unit mass,
\begin{equation}
u_i = \frac{w(r_i,r_0)}{\sum_{j=1}^{n_{\rm [h,k]}} w(r_j,r_0)}\,\frac{E_\star}{m_{\rm g}}\,,
\end{equation}
where $m_{\rm g}$ is the mass of the particle, and $w(r_i,r_0)$ is the kernel weight at distance $r_i$. 
To conserve energy, the kernel weight is normalised to the sum of the weights over all heated/kicked particles.

If the particle is heated, its internal energy increases by $u_i$. If
the particle is kicked, its velocity is increased by $v_i=\sqrt{2
 u_i}$, and the change of kinetic energy in this case equals the
change of thermal energy in the other. The momentum kick is radial,
along the direction from the centre of the explosion to the particle.

For a given number of heated/kicked
particles, because the injected energy per unit mass is
$u_\star\propto E_\star/m_g\propto N$, the mass resolution
defines the energy jump. Our choice of
initial conditions leads to an energy jump of the order of
$\sim10^9$ with respect to the medium initial energy and for the reference value $E_\star=1$.

For given values of $E_\star$ and
$\rho_0$, and assuming monatomic gas with a ratio of specific heats $\gamma=5/3$, the analytic solution for the time evolution
of the shock-front radius is given by \citep{Sedov1959}:
\begin{equation}
R(t) \simeq 1.1527\left(\frac{E_\star}{\rho_0}\right)^{1/5}t^{2/5}\,.
\label{eq:sedovblast}
\end{equation}

Regarding the time integration, we employ in this section global, adaptive time-steps. 
At each simulation step, the whole system is advanced in time by integrating over the minimum time-step of all particles.
To define the time-steps, we use the standard Courant parameter, $C=0.15$, and the value of the time accuracy coefficient $\eta=0.0025$ 
in equations \ref{eq:courdt} and \ref{eq:accdtnew} presented in Appendix~\ref{app:dtcriteria}.
Regarding the accuracy parameter $\eta$, we performed accuracy and efficiency tests with different values, 
and chose $\eta=0.0025$ as fiducial parameter for this study. For a comparison of those tests, 
the reader should refer to Sec.~\ref{sec:sedov} and \ref{sec:evrard}.
Finally, in order to define the SPH  kernel, we set the number of neighbours to $32 \pm 2$ particles.

The list of runs is given in Table~\ref{table:concordstat}. 
In the reference model (highlighted in bold) we adopt the values $E_{\star}=1$ and $\alpha=2$ for the injected energy and the artificial viscosity coefficient, respectively. 
We also run a sample of simulations varying $E_{\star}$ and $\alpha$. 
Beside the numerical parameters of the tests, we also list in Table~\ref{table:concordstat} a measure of the conservation of the injected energy at the end of the runs: 
\begin{equation}
\frac{\Delta E_{\star}}{E_{\star}} = \frac{(E_{+}-E_{0})-E_{\star}}{E_{\star}}\,,
\label{eq:energyconservation}
\end{equation} 
where $E_{+} = E_{\rm t} + E_{\rm k}$ is the time evolution of the sum of the thermal and kinetic energies in the simulated volume 
and $E_0$ is the initial, total energy of the system (excluding the input energy).
We also list the total simulation running time on 8 cores.

\subsection{Simulation results}
\label{subsec:globalsedovresults}

\bigskip

\begin{figure}
  \begin{center}
    \hspace{-2mm}\includegraphics[trim = 0mm 2mm 2mm 4mm, clip, width=0.24\textwidth]{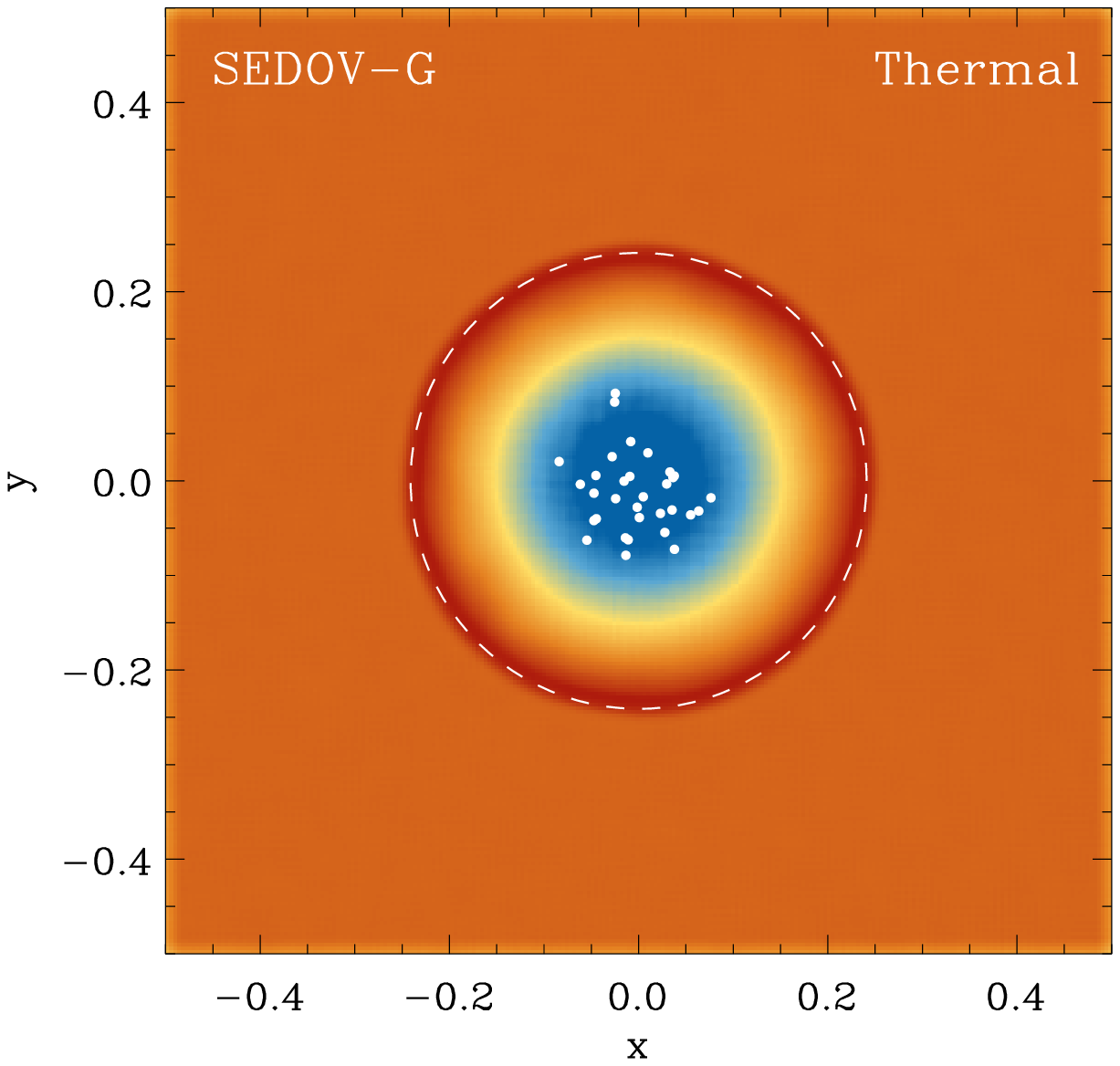}
    \hspace{-1mm}\includegraphics[trim = 0mm 2mm 2mm 4mm, clip, width=0.24\textwidth]{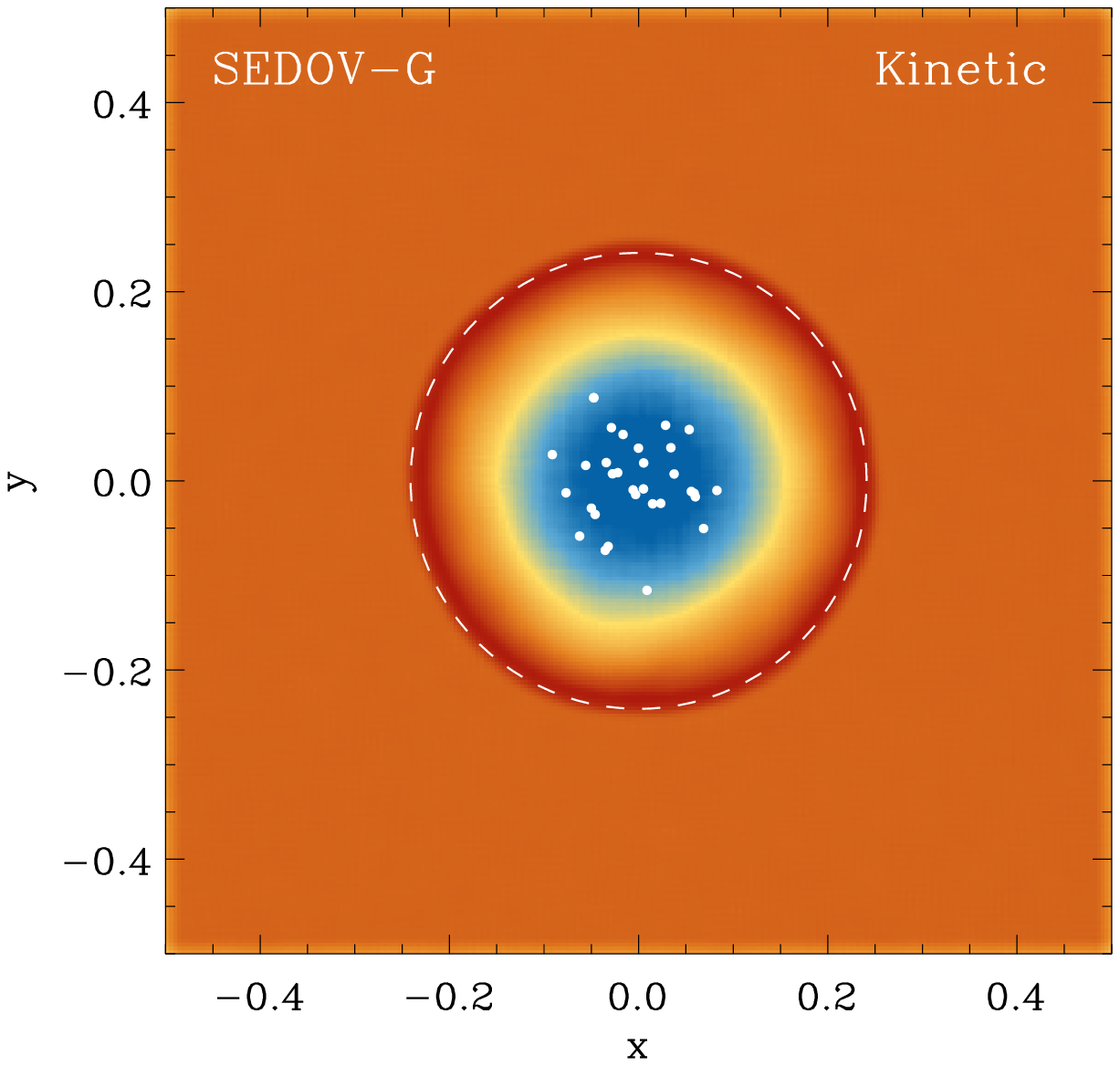} \\
    \hspace{-2mm}\includegraphics[trim = 0mm 4mm 2mm 4mm, clip, width=0.24\textwidth]{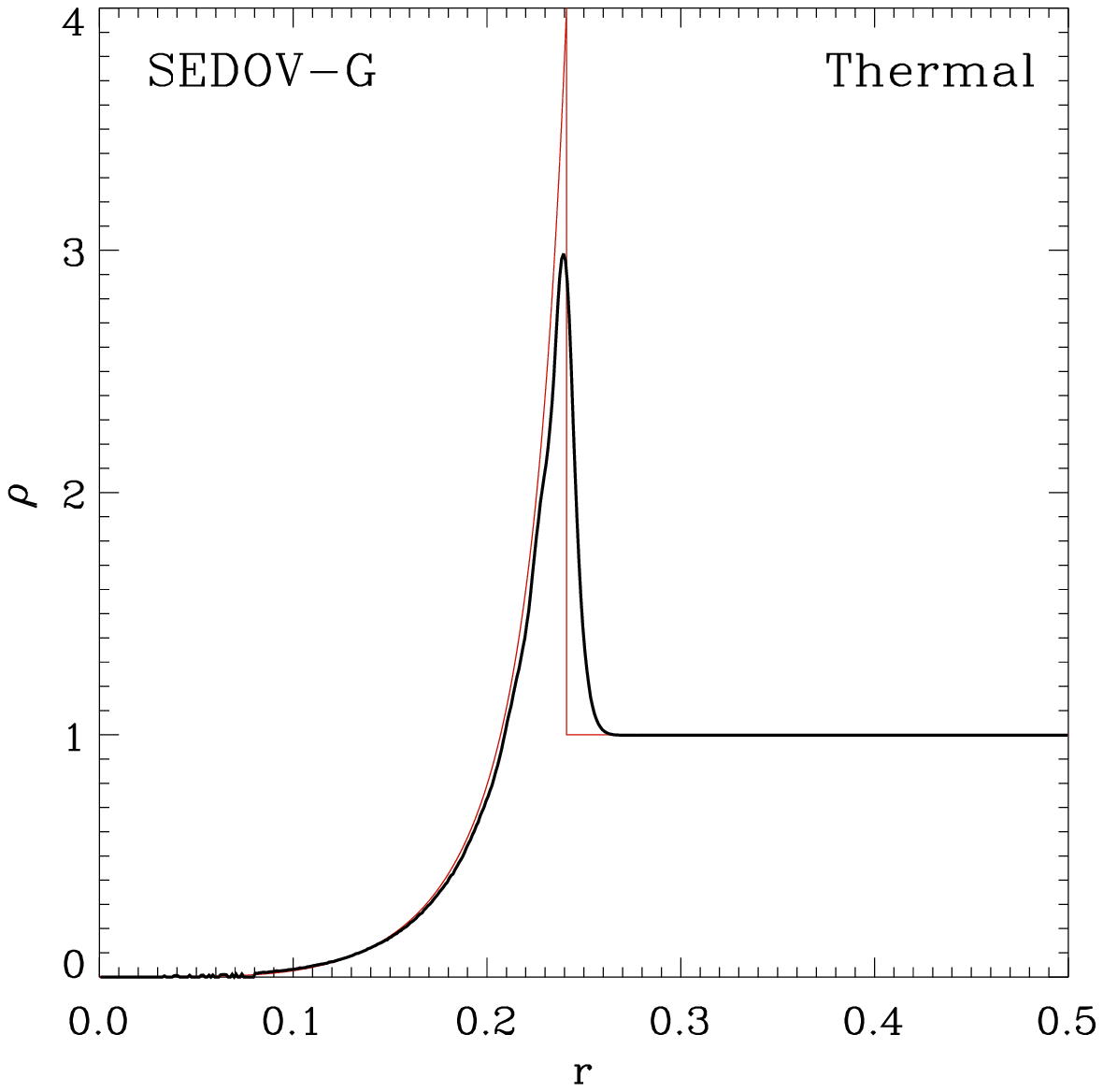}
    \hspace{-1mm}\includegraphics[trim = 0mm 4mm 2mm 4mm, clip, width=0.24\textwidth]{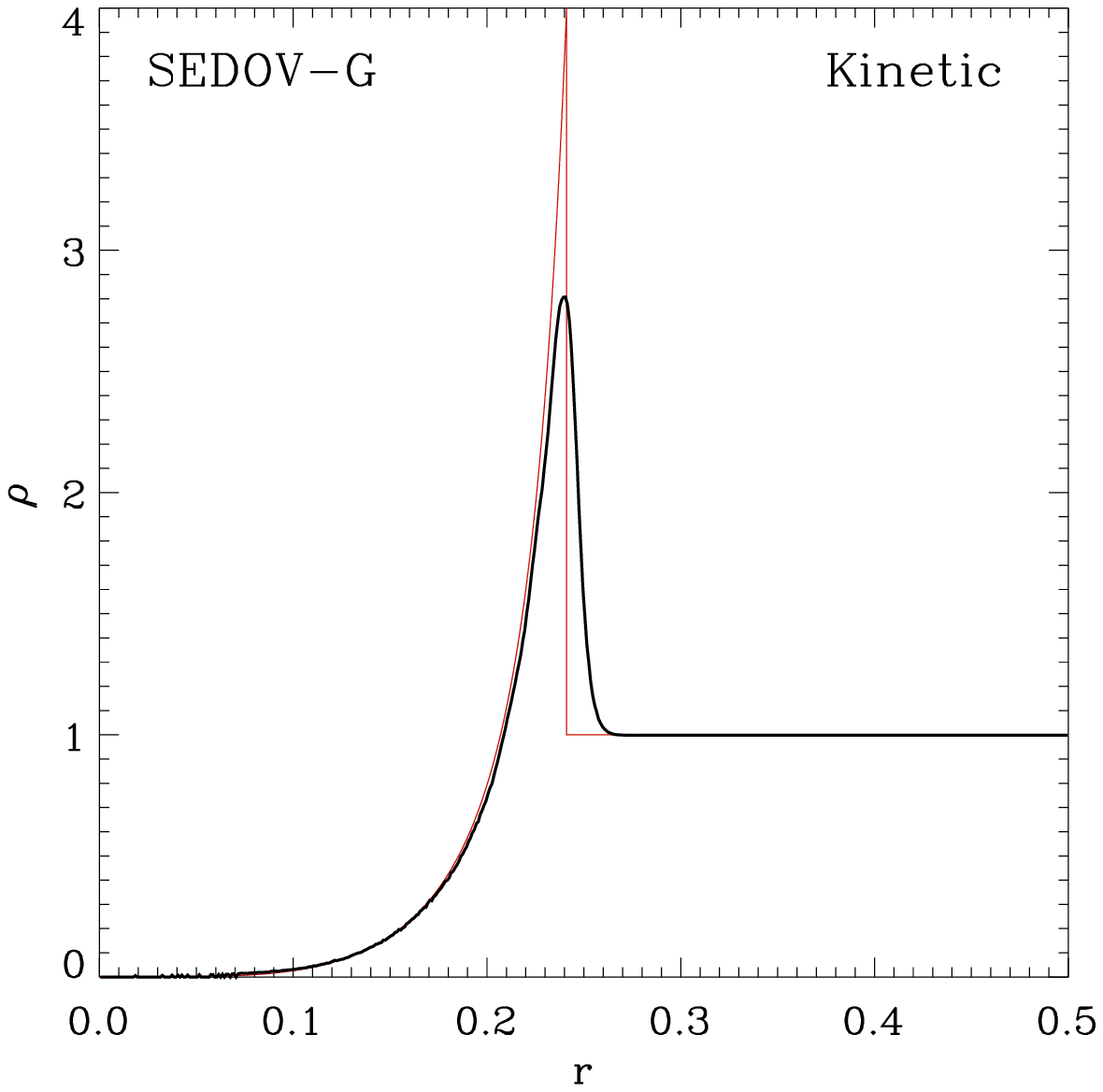}
  \end{center}
  \caption{Sedov's explosion test for the reference simulations, where the system is evolved on global time-steps: 
  energy is injected either in thermal (left column) or kinetic (right column) form.
   Results are given at $t=0.02$ after the explosion.
   We see that both methods agree remarkably well in reproducing the similarity solution for both the location and the width of the blast shell.
    \label{fig:concordprofile}}
\end{figure}

We first describe the reference model by showing in Fig.~\ref{fig:concordprofile} the blast wave behaviour at time $t=0.02$.
In the upper row, the projected gas density of a slice of thickness $\Delta z=0.1$ centred on the box
origin is plotted for the thermal and kinetic feedback runs. 
The white dots mark the position of all particles that initially received the energy. 
The dashed circle corresponds to the position of the spherical shock-front given by the analytic solution in Eq.~\ref{eq:sedovblast} at that time. 
In the lower row, the simulated, average radial density profiles (black lines) are compared against Sedov's similarity solution (red lines).

The agreement with the analytic solution is very good for the two feedback schemes, for both the location and the width of the blast shell. 
As shown by other authors, the lack of resolution and the SPH smoothing do give a density contrast at the shell radius that is smaller than the theoretical prediction
\citep[see for instance a comparative study for AMR and SPH codes by][]{Tasker2008}. 
Nevertheless, we see on the radial profiles that our resolution is sufficient enough to describe correctly the asymmetry of the shock front in both cases. 

As expected from a global time-step integration, the computation time is large, ranging from $\sim3$ to $\sim4$ hours.
With both feedback schemes, the accuracy of energy conservation arises because at each simulation step the entire system is integrated. 
Therefore, all particles are aware of the hydrodynamical state of their neighbours. 
We will develop this argument in more detail in Sec.~\ref{sec:individual}. 
Here, we focus on how and why there is agreement between the two schemes.

It is somehow expected that both methods give the same results. 
Indeed, Sedov's initial conditions are the total energy of the explosion and the medium density. 
The only requirement is that a large amount of energy is instantaneously injected in a small volume,%
\footnote{It has been proved that injecting energy in a region of the size of the spatial resolution gives an accurate treatment of the problem 
\protect\citep[e.g.][]{Tasker2008}.} but its form is not specified \citep{Landau1959}. 
One can input either all thermal, all kinetic or a combination of both forms and obtain the same similarity solution at any given time. 
Since the hydrodynamics conservation laws are used to derive the solution, a property of the similarity solution is that 
the fractions of thermal and kinetic energies of the blast are constant in time.

However, in the numerical integration, a finite time is required to convert one form of energy to the other and reach the energy budget 
given by the analytic solution (see illustration in Fig.~\ref{fig:concordratio}).
As long as momentum can be converted into thermal energy by physical processes like e.g. viscosity,%
\footnote{In SPH simulations kinetic energy is converted to thermal energy by the numerical, artificial viscosity scheme \citep{Monaghan1983,Balsara1995}.} 
the numerical integration of the conservation laws should reach the similarity solution.

\begin{figure}
  \begin{center}
    \hspace{-4mm}\includegraphics[trim = 0mm 4mm 0mm 4mm, clip, width=0.46\textwidth]{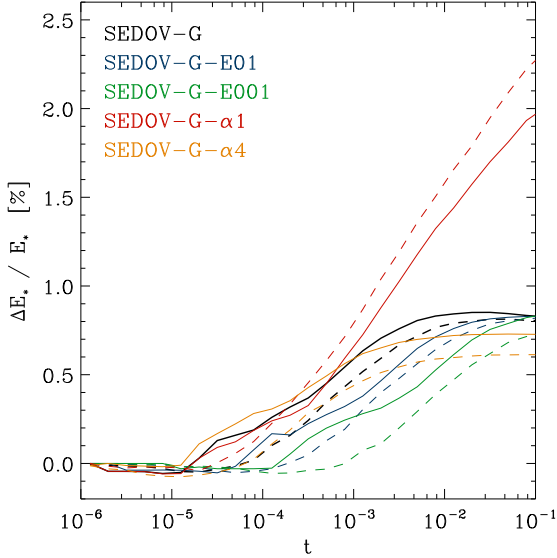}
  \end{center}
  \caption{Time evolution of the input energy conservation for the Sedov's blast wave tests with global time-step integration.
    Solid lines show the results for the thermal injection of energy. Dashed lines represent the evolution of the kinetic feedback runs.
    The colour coding is defined by the properties of the runs.
    \label{fig:concordenergy}}
\end{figure}

\begin{figure}
  \begin{center}
    \hspace{-4mm}\includegraphics[trim = 0mm 4mm 0mm 4mm, clip, width=0.46\textwidth]{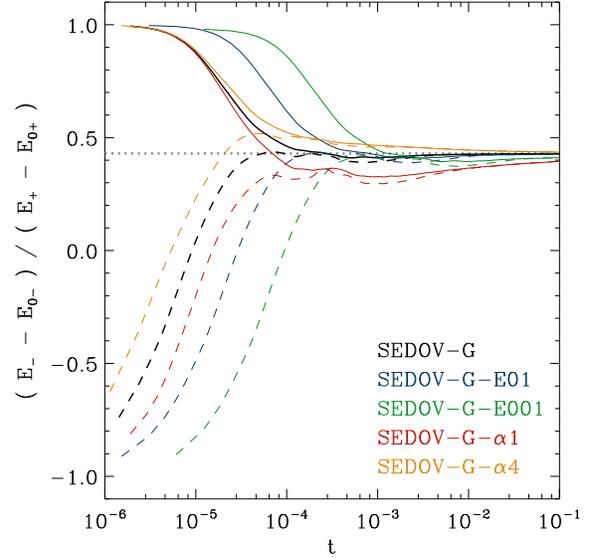}
  \end{center}
  \caption{Time evolution of the energy partition for the Sedov's blast wave tests with global time-step integration.
  	$E_{-}$ and $E_{+}$ are the time evolution of the difference and the sum of 
	the thermal and kinetic budget of the whole system, respectively. 
	Quantities with subscript `0' correspond to the initial values, before energy injection.
    	Solid lines show the results for the thermal injection of energy. 
    	Dashed lines represent the evolution of the kinetic feedback runs. 
    	The colour coding is as in Fig.~{\protect\ref{fig:concordenergy}}.
    \label{fig:concordratio}}
\end{figure}

We show in Fig.~\ref{fig:concordenergy} the time evolution of the energy conservation relative error 
(given by Eq.~\ref{eq:energyconservation}) for all test simulations, 
where solid and dashed lines refer to the thermal and kinetic energy injection methods, respectively. 
Lines of the same colour are for simulations with the same numerical parameters. Energy injection happens at time $t=0$.

We first concentrate on the reference simulation (black lines). 
We show in the plot that the violation of energy conservation happens in the early stage of the explosion ($t\le10^{-2}$), when the energy contrast is the largest. 
In the thermal case (solid line), there is initially a jump of $\sim0.15$\% around $t=10^{-5}$.
The conversion of energy into momentum happens very quickly, and after $t=4\times10^{-4}$ the evolution follows the kinetic case with an offset slowly decreasing. 
At later times, both curves flatten to roughly $0.8$\%.

The same behaviour can be seen in the energy variation tests (blue and green lines), where the input energy is decreased by a factor of 10 and 100, respectively. 
Decreasing the input energy slows the blast evolution and gives the time offset seen the plot. With constant $\alpha$, varying the input energy gives
similar relative errors, providing an estimate that is independent of the energy value. 
Moreover, energy conservation is achieved at a comparable level for both thermal and kinetic feedback.

We show in Fig.~\ref{fig:concordenergy} that the choice of the artificial viscosity parameter plays an important role in the input energy conservation.
We compare three models with $\alpha=1$ (red), $2$ (black) and $4$ (orange).
The simulations with $\alpha=1$ keep an excess of momentum for longer time
(see discussion of Fig.~\ref{fig:concordratio} below), which gives a larger energy conservation error.
On the other hand, simulations with higher $\alpha$ flatten to a smaller energy violation value (below 1\%) at earlier times.
We argue that the error accumulates during the conversion from kinetic to thermal energy in the shock front. 
This indicates that a higher artificial viscosity preserves a better accuracy in shocks.

While the evolution of the input energy relative error informs us on the accuracy of the simulations, 
additional information is needed to understand the behaviour of the blast.
The time evolution of the conversion of one form of energy to the other following the explosion is described in Fig.~\ref{fig:concordratio}. 
In this figure, we show an estimate of the energy partition as $(E_{-}-E_{0-})/(E_{+}-E_{0+})$, where 
$E_{-}=E_{\rm t}-E_{\rm k}$ and $E_{+}=E_{\rm t}+E_{\rm k}$ 
are the time evolution of the difference and the sum of the thermal and kinetic budget of the whole system, respectively. 
In order to follow only the conversion of the input energy $E_{\star}$, we thus remove the initial (before injection) values $E_{0-}$ and $E_{0+}$.
If $E_{\star}$ is in thermal form (solid lines), the ratio is unity at injection time $t=0$; 
if $E_{\star}$ is in kinetic form (dashed lines) the ratio is initially $-1$.

We compare the results with the energy partition expected value given by Sedov's similarity solution. 
At a given time, we calculate the radial profiles of density, velocity and pressure, 
and integrate them numerically to obtain the total thermal and kinetic energy within the blast radius.%
\footnote{Note that in Sedov's solution the total injected energy is within the blast radius.} 
The integration gives 71.7\% of the blast energy in thermal and 28.3\% in kinetic form \citep[see][for the same result]{Chevalier1974}.
We plot the analytic ratio (dotted line) and find agreement with our results.

Indeed, the evolution shows that energy of one form is quickly converted into the other, and that the ratio tends to the same energy partition limit value. 
However, as already mentioned, Fig.~\ref{fig:concordratio} shows clearly that a finite time is needed 
in order for the simulations to converge to the similarity solution. 
We see from the results of the different setups that this convergence time depends on both 
the physical properties of the explosion and the numerical parameters of the simulation.
This property will be studied in more detail in an upcoming work.
We will especially investigate the effect of radiative cooling processes and compare the numerical convergence to the timescale 
at which the adiabatic Sedov's blast evolves to the radiative snowplow phase.

Nevertheless, since decreasing the input energy makes the pressure gradient shallower (thermal feedback) and the viscosity term smaller (kinetic feedback) 
for a given artificial viscosity parameter, the offset seen in the time evolution can already be explained by the lower conversion rates.
However, we see from Fig.~\ref{fig:concordratio} that changing the conversion rate clearly modifies the convergence process. 
With smaller $\alpha$, viscous forces are reduced and more momentum is created in the thermal feedback case.
Conversely, less momentum is dissipated in the kinetic case, even though this excess of momentum from both feedback methods tends, 
slowly with time, to the same expected limit. 
On the opposite, a higher conversion rate ($\alpha=4$) produces an excess of thermal energy, creating too few (thermal case) 
or converting too much (kinetic case) momentum. With this higher $\alpha$ the convergence to Sedov's energy partition happens again slowly with time.
We confirm here that the stability of the numerical scheme depends on the artificial viscosity normalisation.%
\footnote{This supports our claim that the error builds up mostly in the viscous term.}
We also demonstrate that the value of $\alpha=2$ reproduces best Sedov's test through a faster numerical convergence. 
This is in agreement with the result of \cite{Monaghan1992} for the choice of $\alpha$ when considering shock fronts.
However, we cannot yet advise for this somewhat higher value than typically used in cosmological simulations since, 
in the artificial viscosity formalism of \gadgettwo{}, the $\alpha$ parameter does not differentiate shocks from shear flows. 
In the latter, a too high conversion rate would affect dramatically the velocity field at the contact surface.
In this specific matter of the artificial viscosity in SPH simulations, we assess that more work is required before obtaining a more general criterion.

\section{Concordance under efficient time integration}
\label{sec:individual}

\bigskip

\begin{table*}
\begin{center}
\begin{tabular}{ l | l | c | c | c | cc | cc }
\hline
\hline 
\multirow{2}{*}{Name} & \multirow{2}{*}{Scheme} & \multirow{2}{*}{$f_{\rm step}$} & \multirow{2}{*}{$\eta$} & \multirow{2}{*}{$n_{\rm [h,k]}$} & 
\multicolumn{2}{c}{ Energy (\%) } & \multicolumn{2}{c}{ Time (min) } \\ 
& & & & & Thermal & Kinetic & Thermal & Kinetic \\ 
\hline\hline 
{\bf SEDOV-G-D} & {\bf Global - Update} & {\bf --} & {\bf 0.0025} & {\bf 32} & {\bf 0.93} & {\bf 1.35} & {\bf 139} & {\bf 101} \\ 
\hline
SEDOV-L2-NU & Limiter - No Update & 2 & 0.0025 & 32 & $4.0\times 10^4$ & $43.46$ & 119 & 35.4 \\ 
\hline
SEDOV-L2-U & Limiter - Update & 2 & 0.0025 & 32 & 2.37 & 2.19 & 24.6 & 20.8 \\ 
{\bf SEDOV-L4-U} & {\bf Limiter - Update} & {\bf 4} & {\bf 0.0025} & {\bf 32} & {\bf 3.03} & {\bf 2.86} & {\bf 21.8} & {\bf 18.2} \\ 
SEDOV-L8-U & Limiter - Update & 8 & 0.0025 & 32 & 4.85 & 5.28 & 21.0 & 17.7 \\ 
\hline 
SEDOV-L4-U-$\eta$ & Limiter - Update & 4 & 0.025 & 32 & 3.90 & 4.28 & 21.0 & 16.8 \\ 
SEDOV-L4-U-N$4$ & Limiter - Update & 4 & 0.0025 & 4 & 4.12 & 4.33 & 26.4 & 25.7 \\ 
\hline
\end{tabular}
\end{center}
\caption{Energy conservation estimate and computational time for Sedov's blast wave tests with a limiter time-step scheme 
are compared to a reference simulation using global time-steps, when the injection of energy is delayed (D). 
The reference simulations, including our recommended set of parameters when using the limiter, are highlighted in bold characters. 
Names are given to each simulations according to the integration scheme used and when numerical parameters differ from the run of reference.
Energy conservation estimates (in percent) are given at the end of the runs, at time $t=0.04$ after energy injection. The wall-clock time (in minutes) is for runs on 8 processors.
\label{table:sedovstat}}
\end{table*}

In simulations with a large dynamical range (e.g. astrophysical problems of galaxy and structure formation), 
the required time scales can span several orders of magnitude. 
In state-of-the-art cosmological simulations, up to several billions of particles are integrated over time. 
The use of constant and global time integration steps is then prohibitive in terms of computational costs.
The individual time-step integration scheme \citep[first introduced by][]{Aarseth1963,Makino1991} 
allows particles to be integrated on time-steps which are functions of the local state.

SM09 demonstrated that an accurate description of feedback processes which involve strong energy perturbations 
can only be achieved by ensuring fast information transfer when using individual time-steps. 
They proposed an innovative scheme in which the time-step length of neighbouring gas particles is constrained by a limiter.
SM09 validated their time-step limiter algorithm with Sedov's blast wave test.
They also tested the effect of the time-step limiter with a SN-like explosion in a self-gravitating halo of cold gas. 
They showed that the conservation of energy and linear momentum agree remarkably well 
with those obtained with a more conservative (but computationally more expensive) global time-step scheme.

However, SM09 used initial conditions that included the explosion energy. 
This leads to setting the time-step of heated/kicked particles and limiting the time-step of their neighbours at the start of the simulation. 
Therefore, the integration was correctly performed over the appropriate time-step from the beginning. 
In the most general case, e.g. in cosmological simulations where the feedback events from SN or AGN activity do not affect only active particles, 
this would not happen. 
We specifically design our tests to consider the energy input after the simulation has been running for several steps. 
This prevents any time-step adjustment before the integration of the dynamical and hydrodynamical equations is performed. 
We also state here that we do not limit the maximum size of the time-step. 
This setup is justified by the aim of describing the more general case of energetic feedback, 
where individual time-steps reflect only the hydrodynamical state of particles.
In order to study an asymmetric problem, we modify the case of the self-gravitating gas halo by off-setting the explosion. 

In the following sections, we briefly present our method to maintain a high accuracy 
when considering strong feedback events in simulations using an individual time-step scheme.
The reader interested in the implementation of this method should refer to the appendices for a detailed description.
The results of the two sets of test simulations are then presented, focusing on the conditions needed to achieve the concordance between feedback methods.

\subsection{Individual time-stepping for feedback}
\label{subsec:individualscheme}

\bigskip

In the context of the hierarchical time-stepping scheme presented in Appendix~\ref{app:leapfrog}, 
it is important to note that if the energy content of a gas particle is suddenly increased, the particle itself will be informed as soon as it becomes active. 
Within a few active time-steps, a fraction of the energy is efficiently converted from one form to the other (thermal to kinetic or \textit{vice versa}). 
If thermal energy is injected, it will be converted into momentum through strong hydro accelerations due to the large pressure gradient. 
If heated particles and their neighbours do not adjust their time-steps, the integration would lead to very large velocities and an artificial excess of kinetic energy. 
In the other case, the injected kinetic energy is converted into thermal energy through shocks. 
Eventually, the integration over a long time-step would lead to an excess of thermal energy, again violating energy conservation.

In both scenarios, the violation of energy conservation is due to the fact that the particles do not react soon enough to the sudden change of their hydrodynamical state. 
Even if the particle becomes active at the proper time, if the new energy state is not taken into account in defining the length of the next step, 
the integration following the energy increase will also be done over a too long time-step, leading again to non-conservation of energy. 
In any case, it is therefore crucial to capture the initial stage of energy injection. 

We emphasise here the problems that one may encounter when implementing feedback modules in SPH codes.
We show in Appendix~\ref{app:dtcriteria} that the signal velocity and the acceleration are the key information needed to define the time-step. 
In the public release of \gadgettwo{} both quantities are calculated during the computation of the local hydrodynamical forces. 
Consequently, any later change of the energetics of an active particle would not be taken into account during the calculation of the next time-step. 
On the other hand, injecting the energy before the hydro calculation would make the current time-step inconsistent with the new hydrodynamical state of the particle.
If the current time-step is overestimated, the integration could lead to unphysically too strong feedback effects.
We also want to warn the reader against any feedback implementation that would inject the energy as a rate to be applied over a given length of time. 
Modifying either the rate of entropy change \text{(thermal fedback)} or the hydrodynamical acceleration \text{(kinetic fedback)} would not conserve the input energy 
if the kick operator (from the leap-frog integration scheme) is applied on two successive steps of different length.

Considering the problems mentioned above, we recommend injecting instantaneously the feedback energy just before the computation of the next time-step, 
whilst, taking into account the following two actions to preserve the accuracy of the time integration.
Firstly, to ensure the fast information transfer of the change of energy in the medium, we use the time-step limiter proposed by SM09.
In Appendix~\ref{app:timelimiter} we describe an implementation of this limiter in \gadgettwo{} which preserve the time-step synchronisation.
Secondly, to ensure that the computation of the time-step following the explosion takes properly into account the local hydrodynamical change, 
we impose that the particles which receive the energy update their signal velocity and hydrodynamical acceleration. 
Doing so, we impose for these particles an update of the computation of their next time-step 
that will lead to an update of their hydrodynamical properties right after the explosion time 
(see Appendix~\ref{app:timeupdate} for the detail about this time-step update). 
We will now show that these actions are essential to be considered altogether in order to preserve the concordance of feedback methods when using individual time-steps.

\subsection{Sedov's blast wave test}
\label{sec:sedov}


In this section we describe the Sedov's blast wave tests performed with an individual time-step scheme.
Before showing the results of our simulations, we mention the differences with the setup used in Sec.~\ref{subsec:globalsedovsetup}.

\subsubsection{Simulation setup}
\label{subsec:individualsedovsetup}

\smallskip

Starting with the same glass-like uniform conditions presented in Sec.~\ref{subsec:globalsedovsetup}, 
all particles are evolved over global, background time-steps that are of the order of $\Delta t_{\rm  back} \sim 10^{-3}$. 
Since all particles are synchronised from the beginning of the simulation, the issues about an explosion occurring in the middle of active steps 
or about neighbouring particles being initially on different time-bin levels, will not be addressed in this section. 
We refer to Sec.~\ref{sec:evrard} for an analysis of these effects on the long-term evolution of the medium.

To avoid a pre-defined population of the time-bin levels in these Sedov's test simulations, 
the injection of a total energy of $E_\star=1$ (either in thermal or kinetic form), is delayed by a few of the background steps.
In order to focus on feedback processes similar to SN explosions or BH activity, 
we have set the total initial internal energy of the system to $E_0=1$. Given the resolution $N=128^3$ of the simulations, 
the initial energy contrast between the heated/kicked particles and the background ones is of the order of $\sim 10^6$.
All tests presented in this section have been run until $t=0.04$ after the explosion time.

It is interesting to remind here that the amount of injected energy constrains the time-steps that follow the explosion. 
Since the signal velocity \citep[see][]{Monaghan1997} at the explosion location is related to the energy injection as $v_{\rm sig} \propto \sqrt{u_\star}$, 
we can already anticipate through the Courant criterion that, for the energy contrast considered here, 
the step of the heated/kicked particles will be $\sim 10^3$ times smaller than the background step. 
This will correspond to an abrupt drop of about 10 levels in the hierarchy of time-bins.

To analyse the behaviour of our time-step scheme, we use a set of simulations (see Table~\ref{table:sedovstat}) 
where combinations of the integration techniques and numerical parameters are investigated.
Here we use again a simulation with a global time-stepping scheme as reference. 
Then, we test the limiter technique without the time-step update. 
Finally, we enforce the time-step update, as describe in Appendix~\ref{app:timeupdate}. 
With the latter setup, we estimate the impact of the limiter parameter $f_{\rm  step}$, 
as well as the time integration efficiency parameter $\eta$ (from Eq.~\ref{eq:accdtnew}), for both the thermal and kinetic feedback methods. 
With the aim of studying two different initial energy distributions, we also compare simulations
where the energy is injected over a different number of particles.

In order to quantify the accuracy of the different methods, 
we estimate the input energy conservation error at $t=0.04$ for each simulation using Eq.~\ref{eq:energyconservation}.
To estimate the performances of the different time-step schemes, we also list the total running time of each test in Table~\ref{table:sedovstat}.

\begin{figure}
  \begin{center}
    \hspace{-2mm}\includegraphics[trim = 0mm 2mm 2mm 4mm, clip, width=0.24\textwidth]{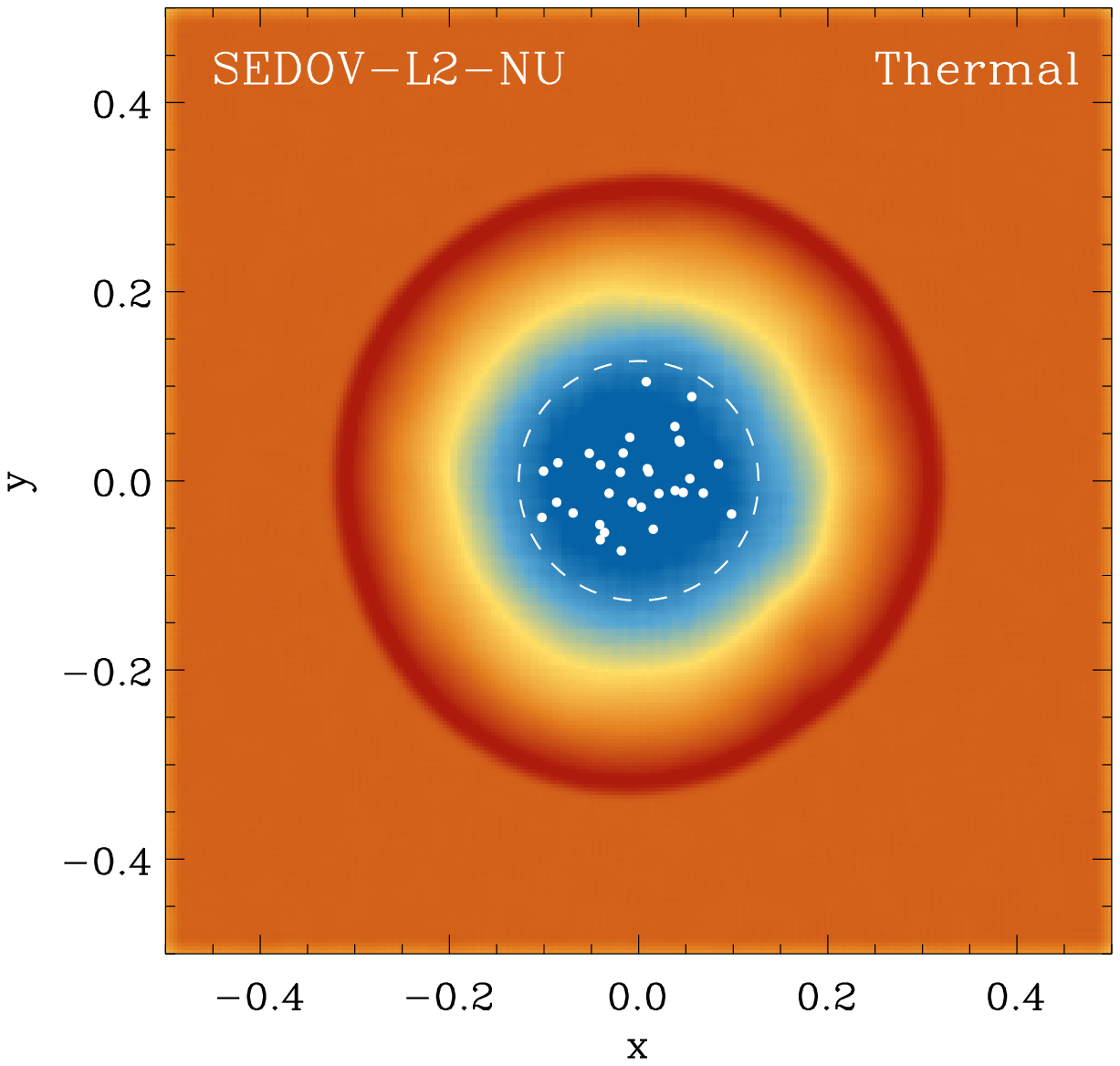}
    \hspace{-1mm}\includegraphics[trim = 0mm 2mm 2mm 4mm, clip, width=0.24\textwidth]{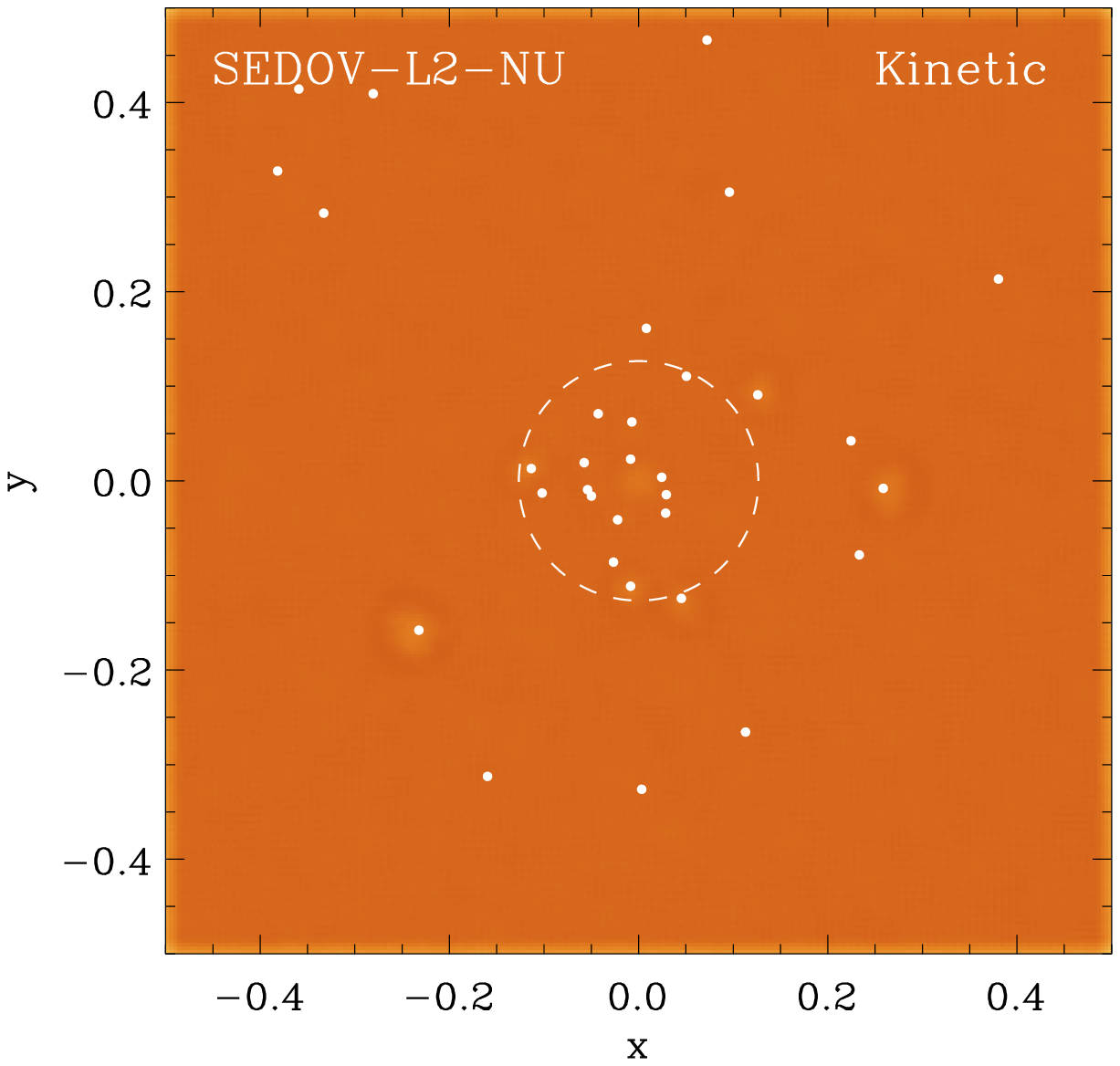} \\
    \hspace{-2mm}\includegraphics[trim = 0mm 4mm 2mm 4mm, clip, width=0.24\textwidth]{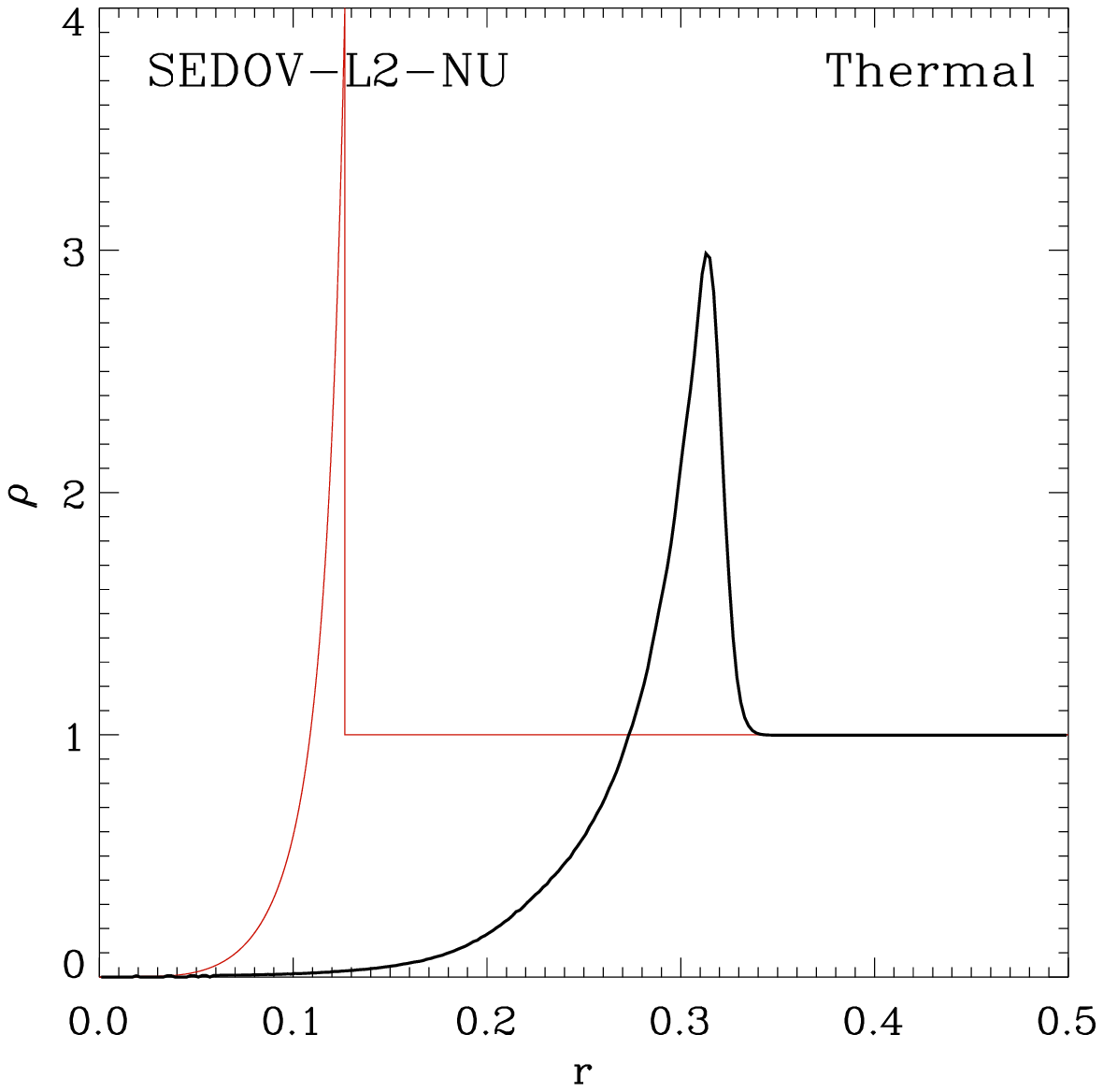}
    \hspace{-1mm}\includegraphics[trim = 0mm 4mm 2mm 4mm, clip, width=0.24\textwidth]{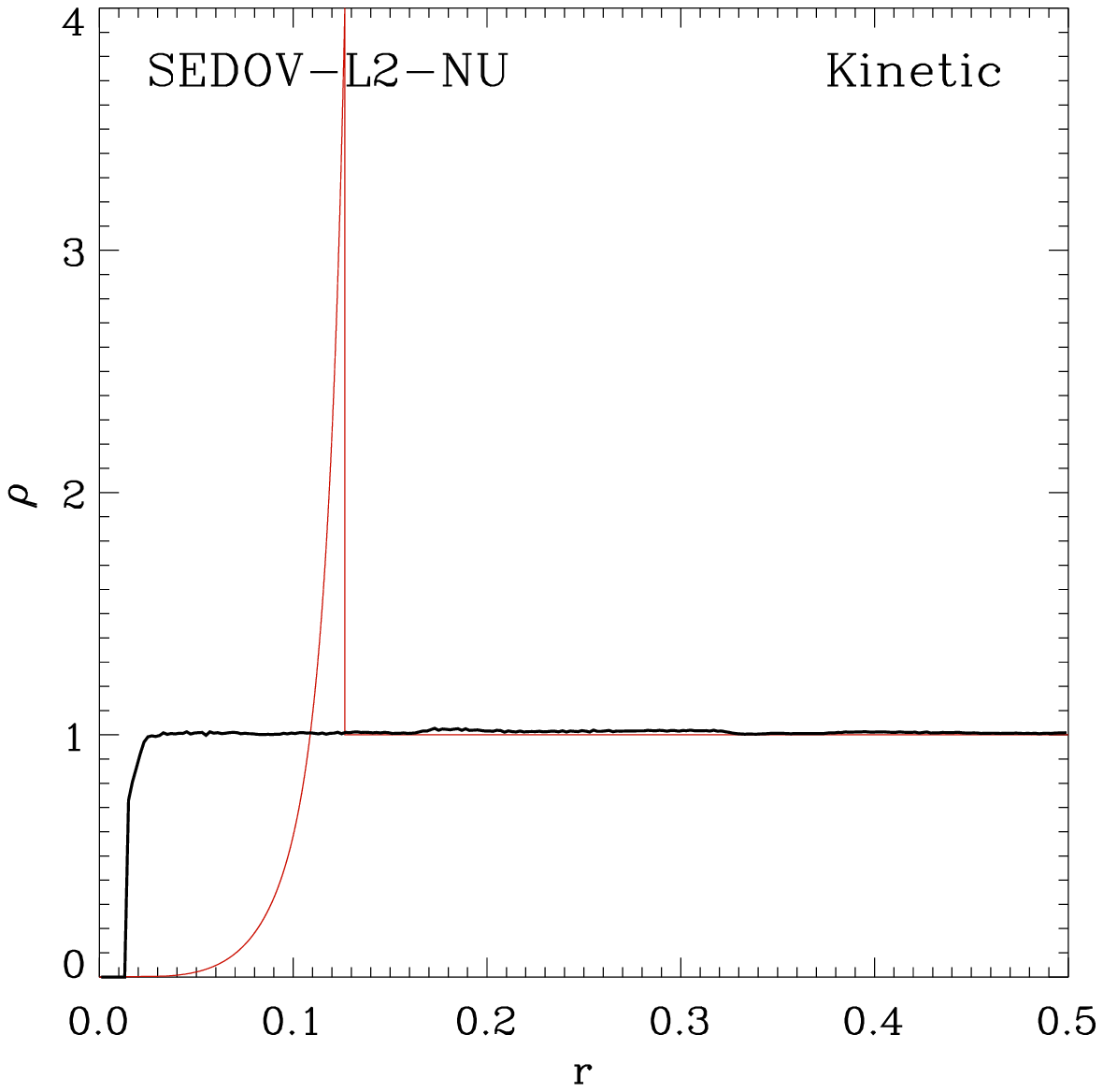}
  \end{center}
  \caption{Sedov's explosion test for thermal (left column) and kinetic (right column) energy injection, 
  when the limiter is applied without the time-step update. Results are given soon after the explosion at time $t=0.004$.
  We see that for the thermal implementation of feedback a stable shell has developed far ahead from the similarity solution 
  because the large violation of energy conservation largely increases the system total energy (see Table~\ref{table:sedovstat}). 
  In the kinetic feedback case inter-particle crossing is evident. 
  No density contrast has yet developed given that kicked particles have traveled far from the explosion site. 
  These results illustrate how the lack of a prompt response of the medium leads to two distinct effects from the two feedback schemes.
  \textit{Time animation available as supporting online material.}
  \label{fig:sedovnocorrect}}
\end{figure}

\begin{figure}
  \begin{center}
    \hspace{-2mm}\includegraphics[trim = 0mm 2mm 2mm 4mm, clip, width=0.24\textwidth]{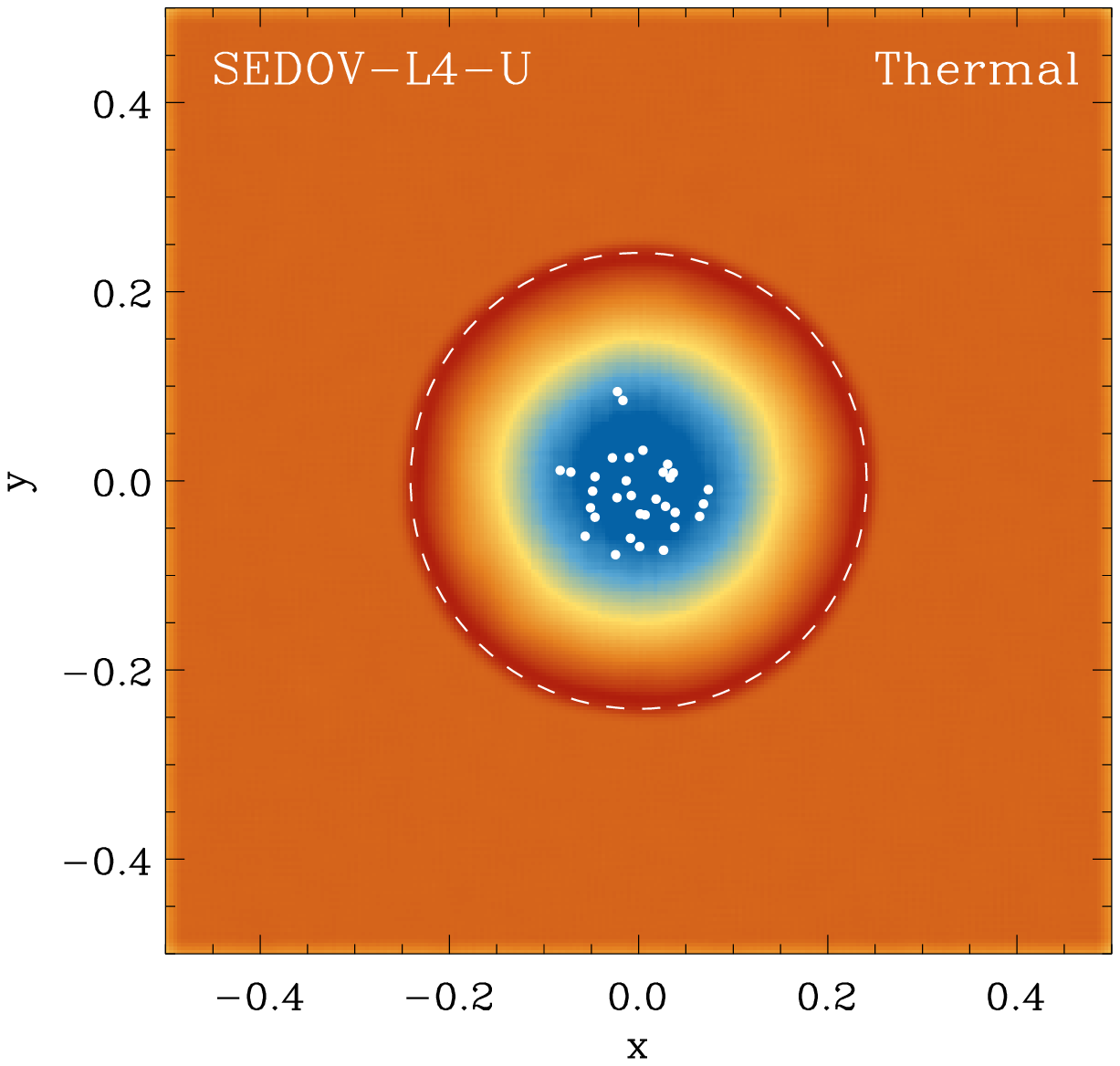}
    \hspace{-1mm}\includegraphics[trim = 0mm 2mm 2mm 4mm, clip, width=0.24\textwidth]{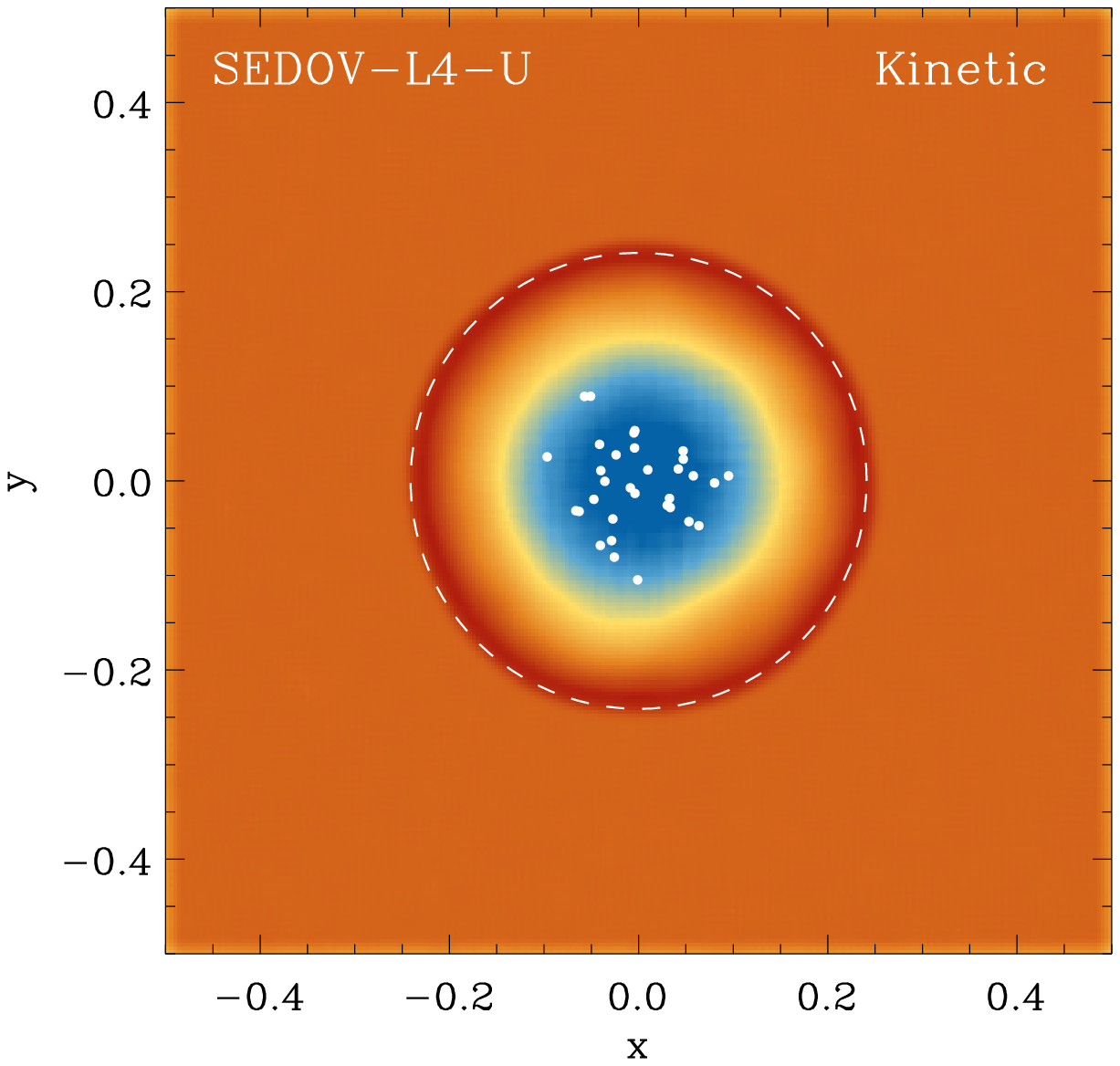} \\
    \hspace{-2mm}\includegraphics[trim = 0mm 4mm 2mm 4mm, clip, width=0.24\textwidth]{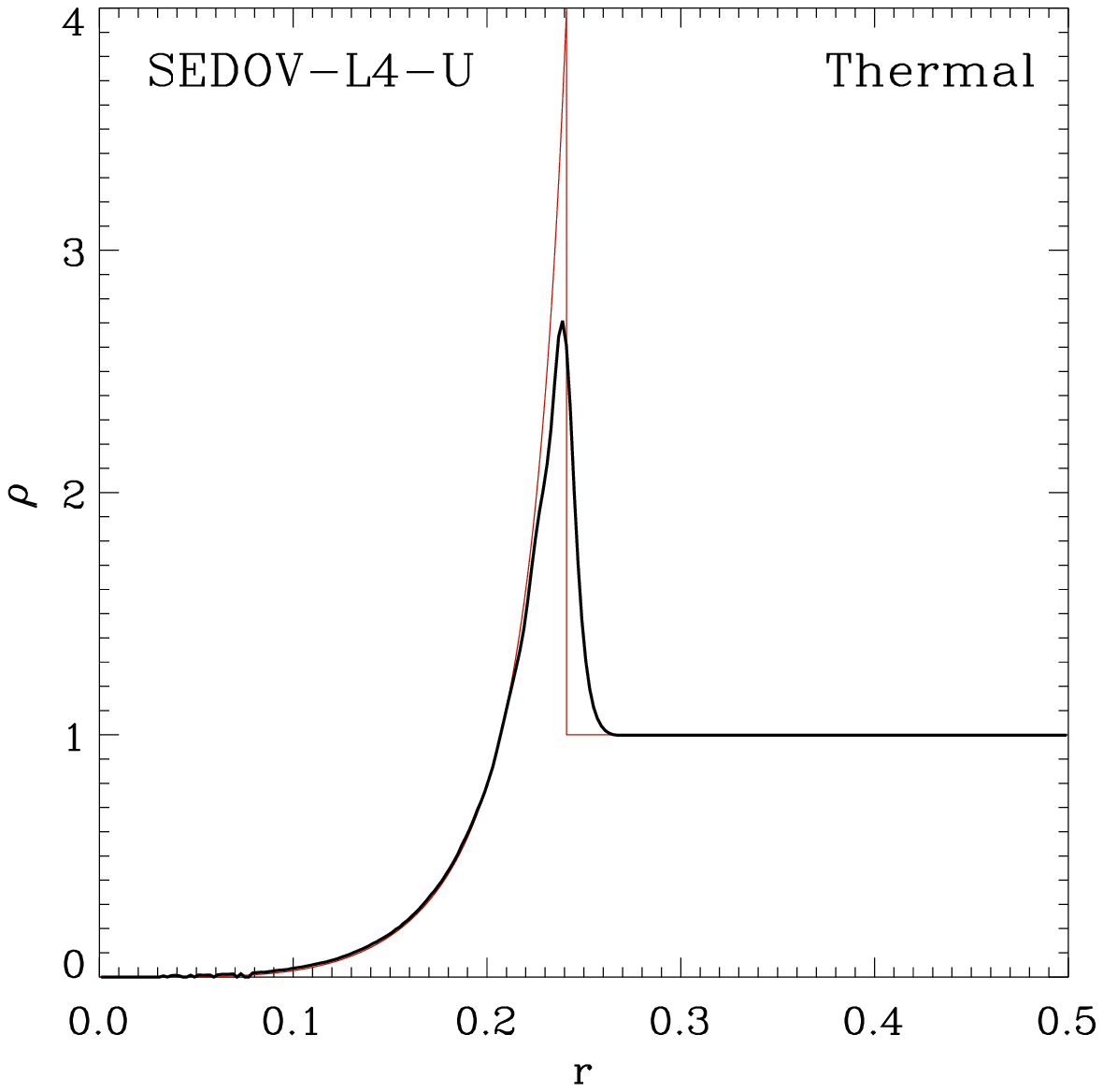}
    \hspace{-1mm}\includegraphics[trim = 0mm 4mm 2mm 4mm, clip, width=0.24\textwidth]{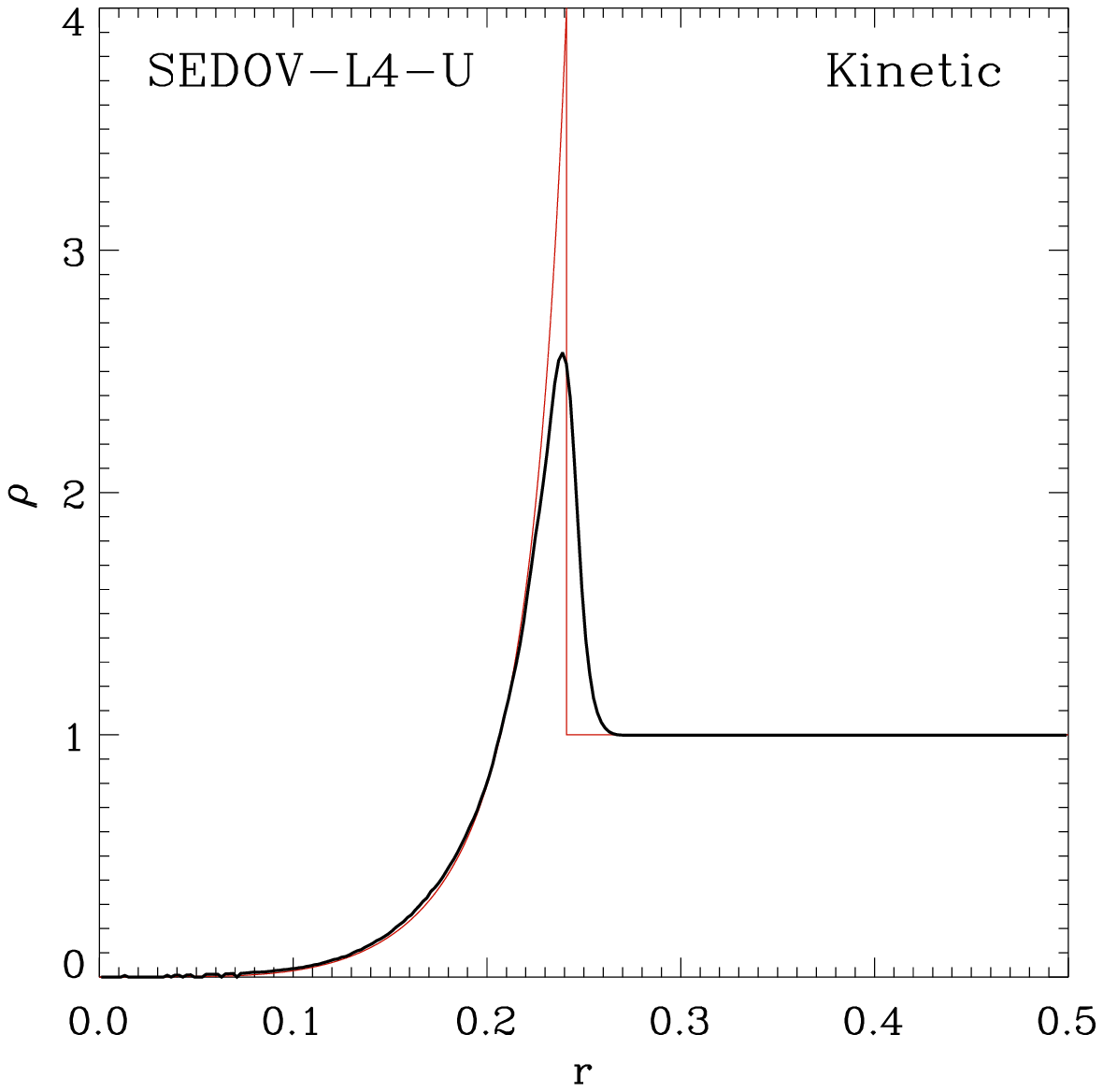}
  \end{center} 
  \caption{Sedov's explosion test for thermal (left column) and kinetic (right column) energy injection, 
  when both the limiter and the time-step update are applied. Compared with the previous figure, results are shown at later time $t=0.02$.
  We show here that, for our recommended set of parameters, the two feedback schemes are able to reproduce precisely the similarity solution 
  and hence provide again concordant results as in Fig.~\ref{fig:concordprofile}.
  This demonstrate that both a fast information transport and a prompt response to the explosion are needed in order to resolve strong feedback events 
  with individual time-step integration.
  \textit{Time animation available as supporting online material.}
    \label{fig:sedovcorrect}}
\end{figure}

\subsubsection{Simulation results}
\label{subsec:sedovresults}

\smallskip

We present in Fig.~\ref{fig:sedovnocorrect} the early state of the medium at $t=0.004$, for both thermal (left column) and kinetic (right column) energy injection, 
when using the time-step limiter without any update at the explosion time. 
This corresponds to the time integration that has been illustrated in the appendices in Fig.~\ref{fig:timelines}-b). 
For those runs, we have chosen the conservative value of $f_{\rm step}=2$ in order to ensure the fastest information propagation between neighbouring particles. 
It is striking to see how the two feedback methods give extremely deviant but very different results. 
In the thermal case, heated particles are found close to the analytical blast radius but since the first step after the explosion is much too long, 
the acceleration given to the surrounding medium produces an important excess of kinetic energy. 
Eventually, the propagation of this excess of energy will lead to a violation of energy conservation to a level of $\sim 40,000\:\%$ at the end of the run.

For the kinetic injection of energy, we see that the isotropic nature of the blast is destroyed. 
Indeed, the lack of an immediate response of the medium makes kicked-particles cross their neighbours 
before they partially thermalise their momentum. Later on, they will eventually start to expand individual bubbles around them. 
Even with the use of the limiter, some particles are able to travel up to the border of the volume before interacting with the medium. 
Here, the energy creation reaches a level of $\sim 43\:\%$. 
This two pictures clearly show the importance of properly computing the time-step following the energy injection event.

In Fig.~\ref{fig:sedovcorrect}, we show the results for the simulations in which we additionally enforce the time-step update. 
Here we use the fiducial value of $f_{\rm step}=4$. The shell position and the radial density profile are in extremely good agreement with the analytic solution. 
Sedov's solution is remarkably well represented for both feedback methods, reproducing again the concordance 
with a conservation of input energy close to $3\,\%$ at the end of the runs.

\begin{table*}
  \begin{center}
    \begin{tabular}{ l | l | c | c | c | cc | cc }
      \hline\hline 
      \multirow{2}{*}{Name} & \multirow{2}{*}{Scheme} & \multirow{2}{*}{$f_{\rm step}$} & \multirow{2}{*}{$\eta$} & \multirow{2}{*}{$n_{\rm [h,k]}$} & 
      \multicolumn{2}{c}{ Energy (\%) } & \multicolumn{2}{c}{ Time (h) } \\ 
      & & & & & Thermal & Kinetic & Thermal & Kinetic \\ 
      \hline 
      \hline 
          {\bf HALO-G-D} & {\bf Global - Update} & {\bf --} & {\bf 0.0025} & {\bf 32} & {\bf 1.62} & {\bf 1.84} & {\bf 25.70} & {\bf 25.87} \\ 
          \hline 
          HALO-I-U & Individual - Update & -- & 0.0025 & 32 & $5.7\times 10^4$ & $1.7\times 10^5$ & 14.20 & 19.73 \\ 
          HALO-L2-NU & Limiter - No Update & 2 & 0.0025 & 32 & $3.0\times 10^3$ & 19.42 & 6.07 & 3.73 \\ 
         \hline 
          HALO-L2-U & Limiter - Update & 2 & 0.0025 & 32 & 2.01 & 1.27 & 3.63 & 3.60 \\ 
          {\bf HALO-L4-U} & {\bf Limiter - Update} & {\bf 4} & {\bf 0.0025} & {\bf 32} & {\bf 2.16} & {\bf 2.60} & {\bf 3.41} & {\bf 3.39} \\ 
          HALO-L8-U & Limiter - Update & 8 & 0.0025 & 32 & 2.27 & 3.26 & 3.35 & 3.34 \\ 
          \hline 
          HALO-L4-U-$\eta$ & Limiter - Update & 4 & 0.025 & 32 & 4.05 & 5.51 & 2.40 & 2.38 \\ 
          HALO-L4-U-N$1$ & Limiter - Update & 4 & 0.0025 & 1 & 2.60 & 0.95 & 3.44 & 3.47 \\ 
          \hline
    \end{tabular}
  \end{center}
  \caption{Energy conservation estimate and computational time for the off-centre explosion tests, where the energy injection is delayed (D) after the start of the simulation. 
The reference simulations, including our recommended set of parameters when using the limiter, are highlighted in bold characters. 
Names are given to each simulations according to the integration scheme used and when numerical parameters differ from the run of reference.
Energy conservation estimates (in percent) are given at the end of the runs, at time $t=0.04$ after energy injection. The wall-clock time (in hours) is for runs on 8 processors.
 \label{table:evrardstat}}
\end{table*}

Regarding the parameter study shown in Table~\ref{table:sedovstat}, we can see that increasing the value of $f_{\rm step}$ slowly enhances the energy violation, 
while keeping the computational time nearly constant. Using a less conservative accuracy parameter $\eta=0.025$ increases also the energy creation 
but both thermal and kinetic simulations are completed faster. 
We note that even distributing the energy over a smaller number of neighbours, which provides a larger energy contrast, gives an acceptable energy conservation.

\begin{figure}
  \begin{center}
    \hspace{-4mm}\includegraphics[trim = 0mm 4mm 0mm 4mm, clip, width=0.46\textwidth]{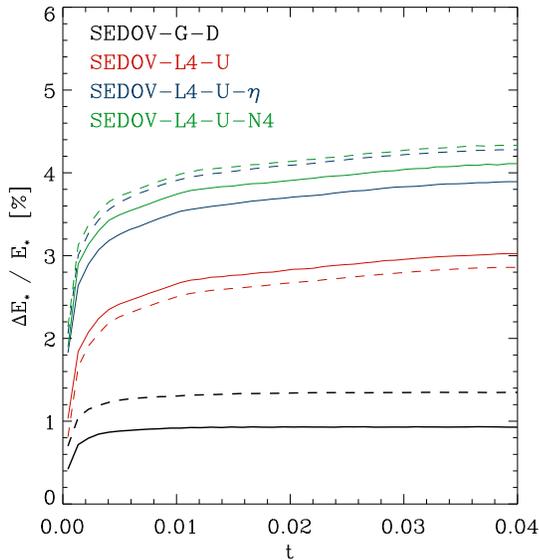}
  \end{center}
  \caption{Time evolution of the energy conservation for the most relevant Sedov's blast wave tests. 
    Solid lines show the results for the thermal injection of energy. 
    Dashed lines represent the evolution of the kinetic feedback runs.
    \label{fig:sedovenergy}}
\end{figure}

In Fig.~\ref{fig:sedovenergy}, the time evolution of the energy
conservation for the runs using the fiducial value of $f_{\rm step}=4$
is compared with the reference simulations using the global time-step
scheme.  We notice that the ratio of energy increase between matching
models is fairly constant.  This behaviour tells us that the
convergence between the two feedback methods is achieved since
the very beginning of the expansion of the blast.  Moreover, we see in
this figure that all simulations experience first a sharp increase
of the energy creation before reaching a stable level of energy
conservation.  Indeed, the rate of energy violation is high right
after the explosion (when the energy needs to be absorbed by the
surrounding of the injection region) but as soon as the blast becomes
stable and starts to expand, the propagation of the error becomes very
small. This happens approximately for $t>0.004$.  We also confirm from
this analysis that the time accuracy parameter $\eta$ needs to be
chosen small enough in order to get the best convergence with the
feedback methods.  Finally, regarding the spatial energy distribution,
it is clear that concentrating the injection over a smaller number of
particles gives a higher energy jump (meaning a larger time-level
drop), and hence produces a slightly worse energy conservation.

We conclude that, when applying the proposed integration scheme, 
there are no dramatic differences either in concordance, or in energy conservation accuracy, 
or in performances for the set of parameters we explored.

\subsection{A more realistic test}
\label{sec:evrard}

\bigskip

In this section we simulate an explosion event in a self-gravitating
gas halo. The initial conditions are similar to the ones in SM09.
However, the injected energy is placed off-centre to follow the blast
evolution in the presence of pressure and density gradients.  In
cosmological simulations of structure formation, the sources of
mechanical and thermal feedback (SNe and BHs) are generally situated
in a similar environment. Though idealised, the setup gives a
qualitative and quantitative view of the above scenario.
                
\subsubsection{Simulation setup}
\label{subsec:evrardsetup}

\smallskip

We create a spherical particle distribution of density profile
$\rho\propto r^{-2}$ by spatially remapping a uniform, spherical
distribution of particles through the radial transformation
\begin{equation}
r_{{\rm old},i}\rightarrow r_{{\rm new},i}=\left(\frac{r_{{\rm old},i}}{R}\right)^3 R\,,
\end{equation}
where $R=1$ in the same system of units used for Sedov's blast wave
problem.  The sphere is cut out from a uniform distribution of
$N=128^3$ particles in a cubic volume of side $L=2$.  The total mass
of the gas sphere is unity. We assign an internal energy of 0.05.  The
system is evolved including self-gravity up to time $t=3$ when relaxation has already
taken place \citep{Evrard1988}.  We chose the gravitational softening
$\epsilon=3.125\times 10^{-3}$ and the artificial viscosity parameter
$\alpha=1$.

We start the simulation tests using the relaxed gas halo as initial conditions, and inject either thermal or kinetic energy in a region centred on $[0.05,0,0]$. 
The off-centre explosion is generated by enhancing the energy of particles by $u_{\rm [h,k]}$ as described in Sec.~\ref{subsec:globalsedovsetup}.  
We avoid the first simulation step in which all particles are active, and delay the energy injection to the time $t_{\star}=10^{-4}$. 
For the sake of simplicity, in the whole analysis, we refer to the explosion time as the initial time $t_0=t_{\star}$. 
Any other time is relative to $t_0$. To remove another bias, we also ensure that at injection time all particles receiving energy are inactive.
Doing so, we can apply our time-stepping scheme to the most generic situation, but more importantly, put it also under the least favourable conditions.

\begin{figure}
  \begin{center}
    \hspace{-.35cm} \includegraphics[trim = 0mm 8mm 4mm 8mm, clip, width=0.245\textwidth]{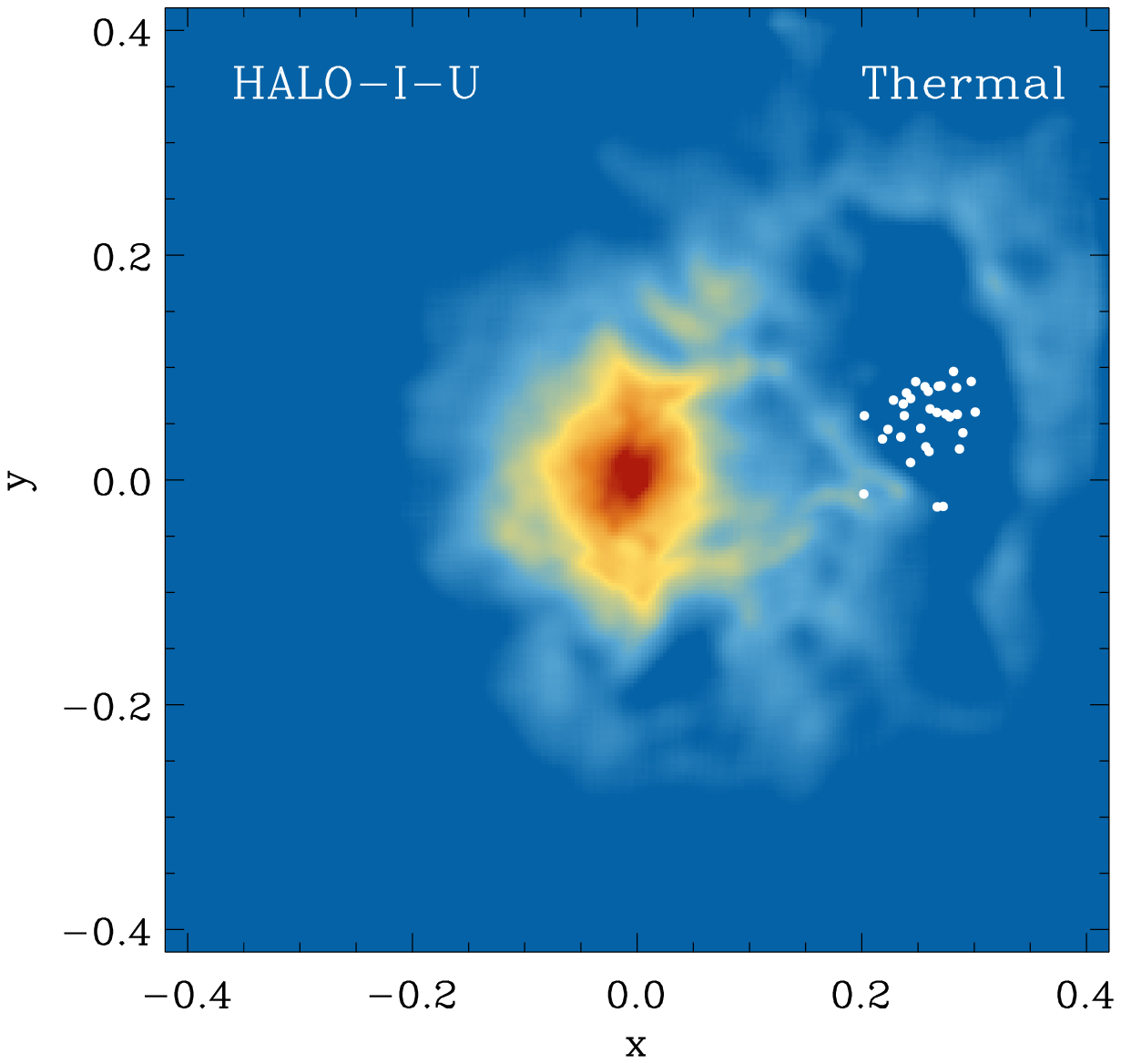}
    \includegraphics[trim = 8mm 8mm 4mm 8mm, clip, width=0.23\textwidth]{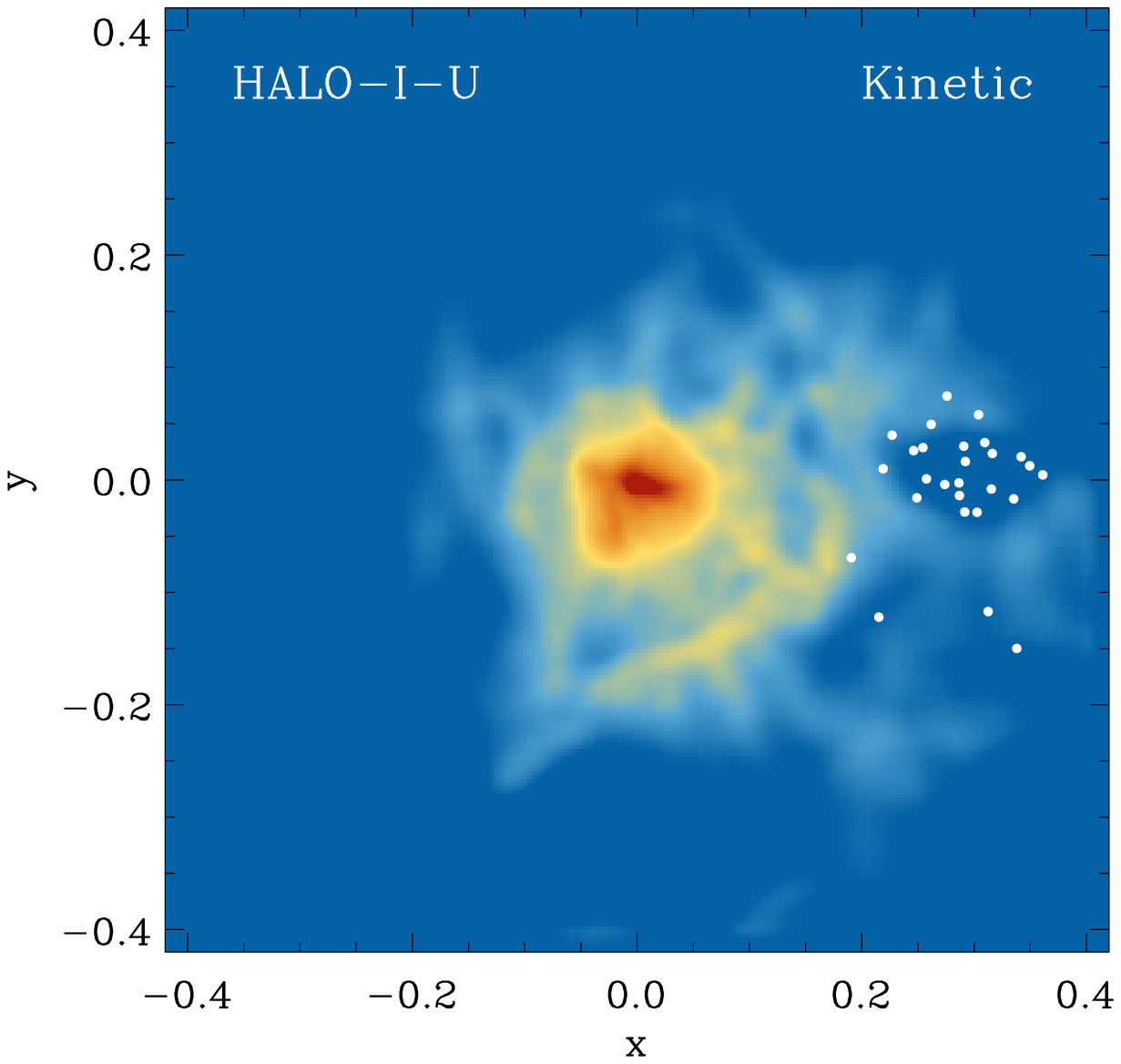}\\
    \hspace{-.35cm} \includegraphics[trim = 0mm 0mm 4mm 8mm, clip, width=0.245\textwidth]{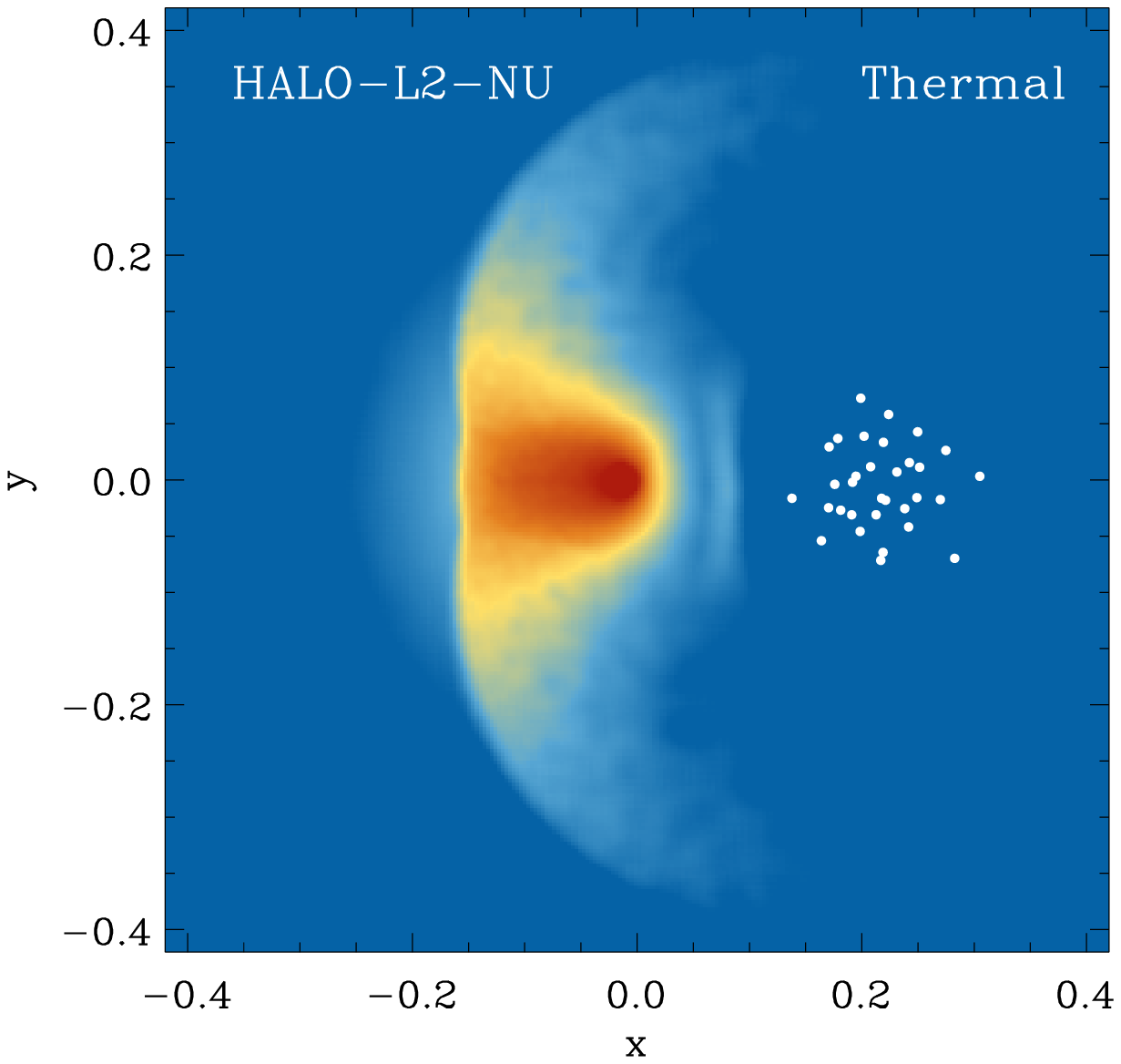}
    \includegraphics[trim = 8mm 0mm 4mm 8mm, clip, width=0.23\textwidth]{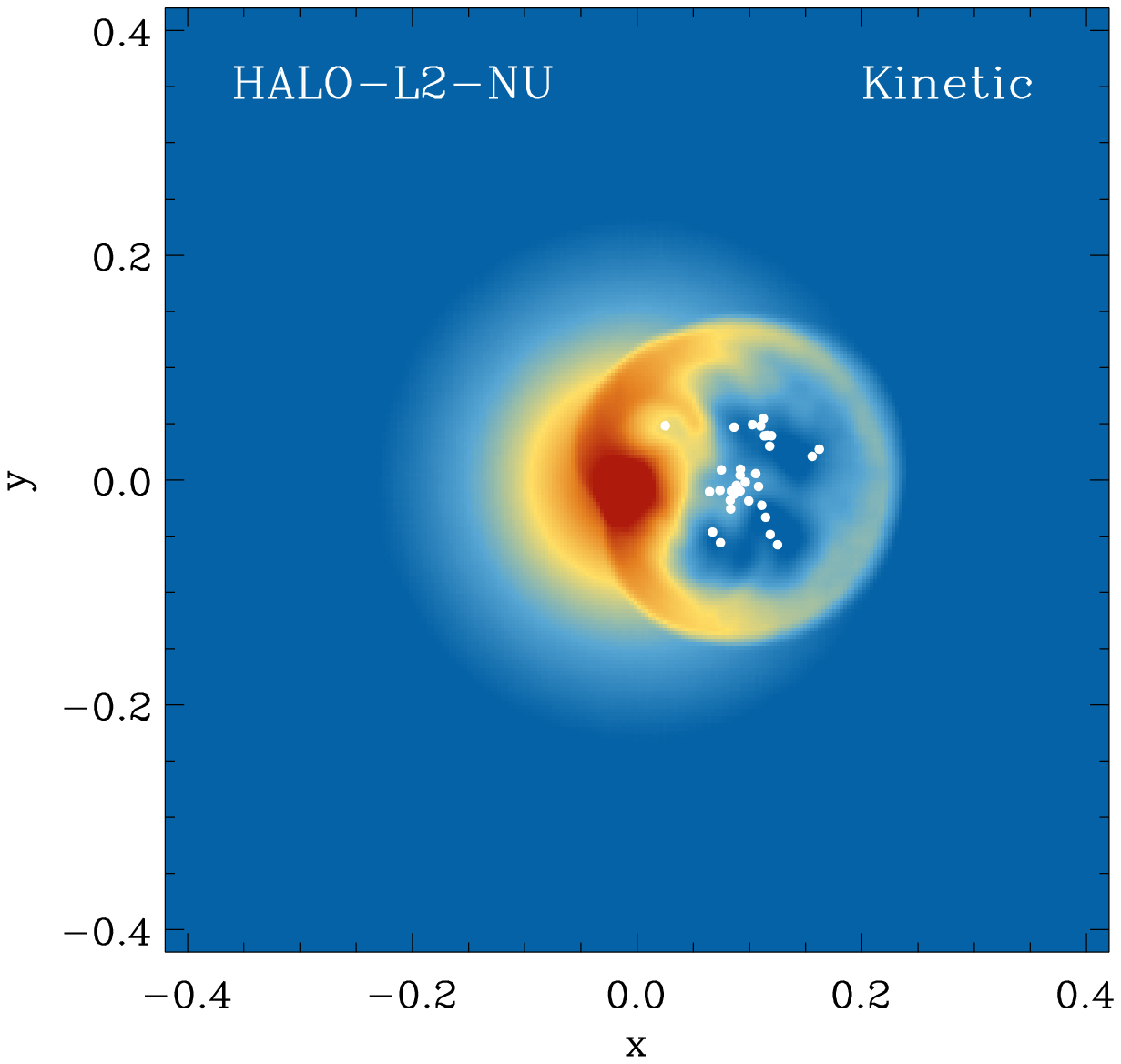}
  \end{center}
  \caption{The off-centre explosion in a self-gravitating gas sphere for thermal (left column) and kinetic (right column) energy injection.
      In all pictures we show the gas density at time $t=0.02$ in a slice of thickness $\Delta z=0.1$ centred on the box origin.
      The white dots mark the position (at the same time) of all particles that received thermal or kinetic energy.  
      \textit{Top row:} only the time-step limiter is applied using the conservative value of $f{\rm step}=2$. 
      The energy violation is severe in the thermal case and the bubble have blown away a large fraction of the gas halo. 
      \textit{Bottom row:} only the time-step update is enforced. The behaviour is similarly wrong in both cases. 
      The halo atmosphere is disrupted and no expanding bubble forms.
      \textit{Time animation available as supporting online material.}
      \label{fig:evrwrong}}
\end{figure}

\begin{figure}
  \begin{center}
    \hspace{-.3cm} \includegraphics[trim = 0mm 8mm 4mm 8mm, clip, width=0.245\textwidth]{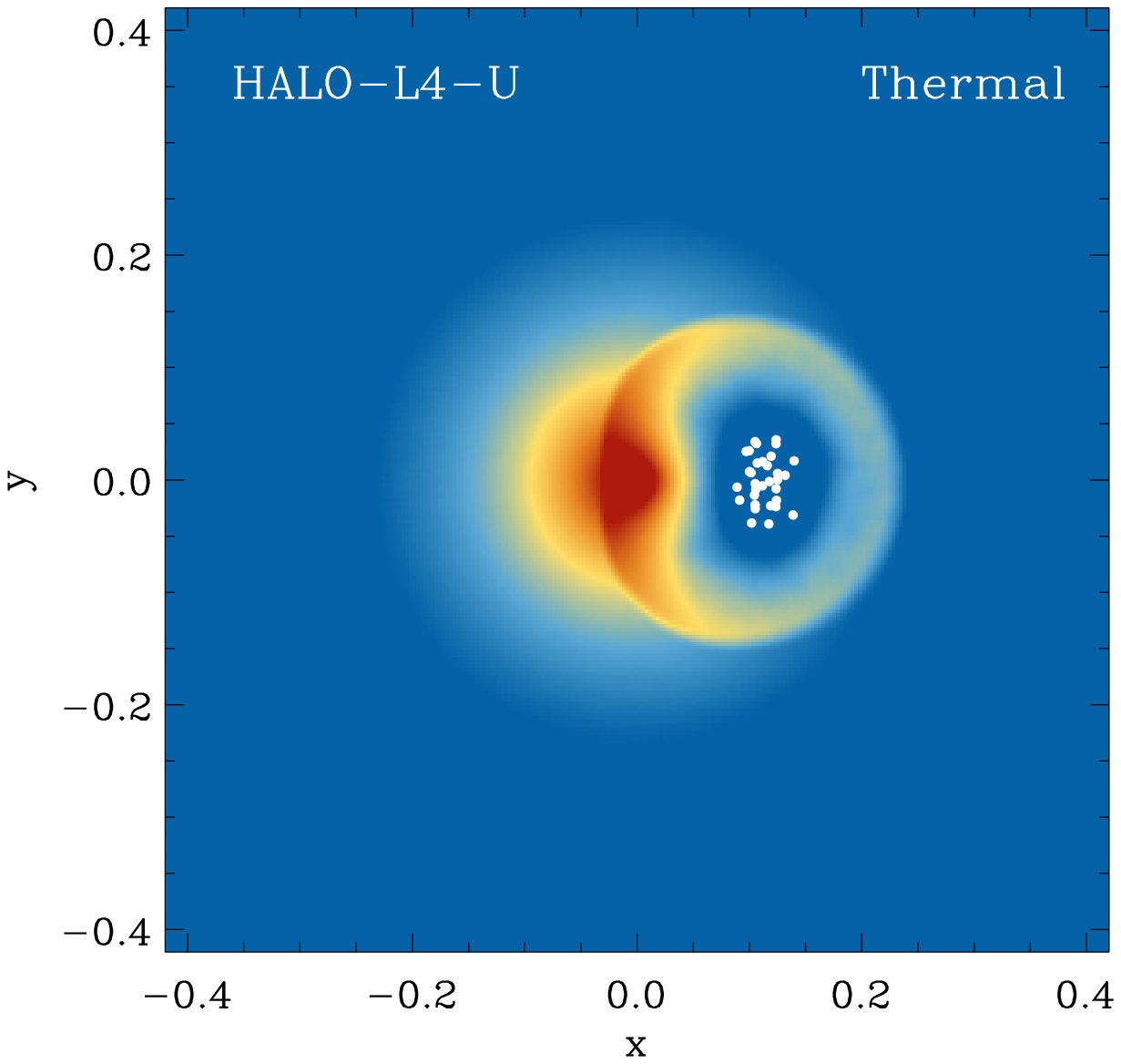}
    \includegraphics[trim = 8mm 8mm 4mm 8mm, clip, width=0.23\textwidth]{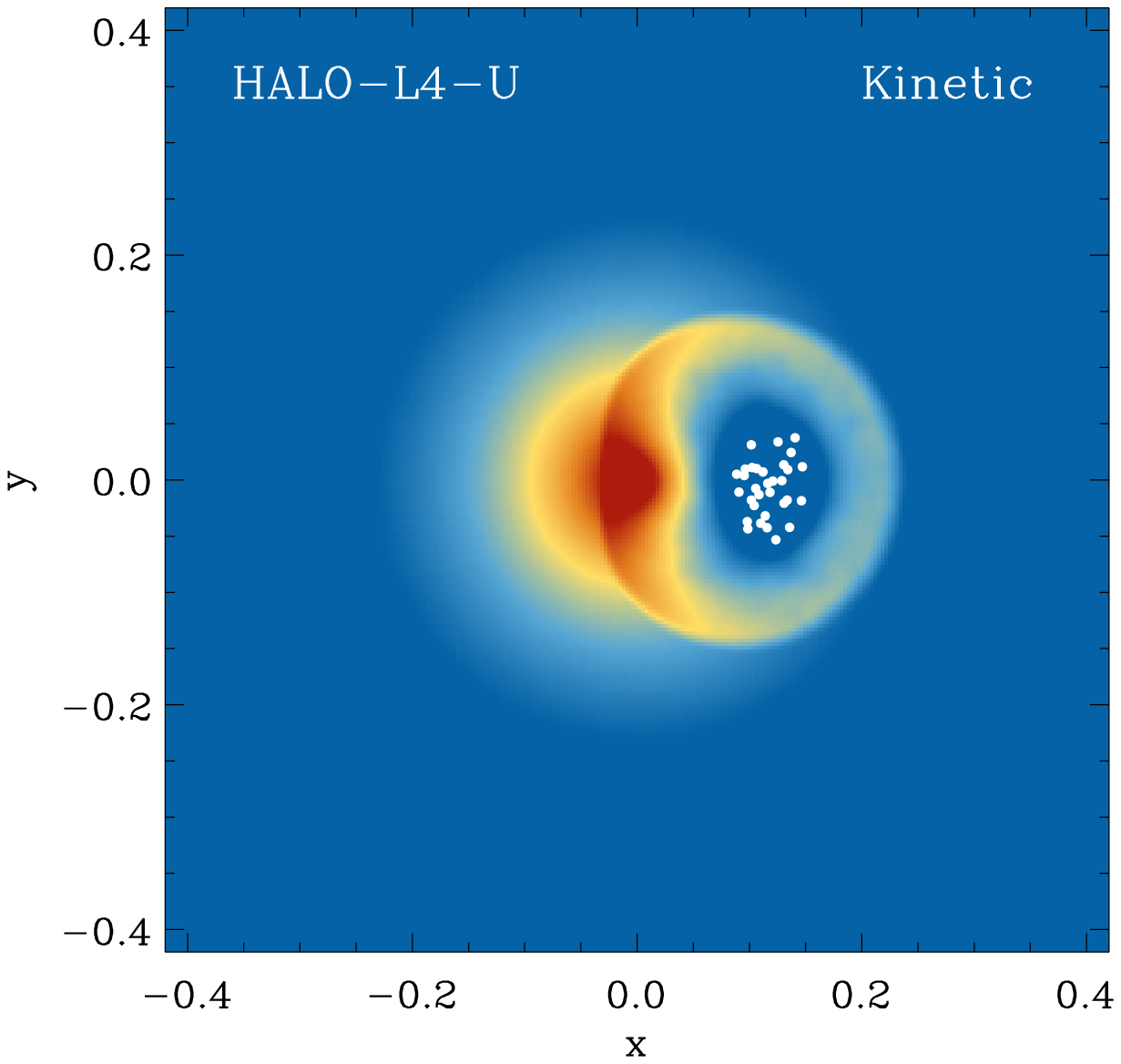}\\
    \hspace{-.3cm} \includegraphics[trim = 0mm 0mm 4mm 8mm, clip, width=0.245\textwidth]{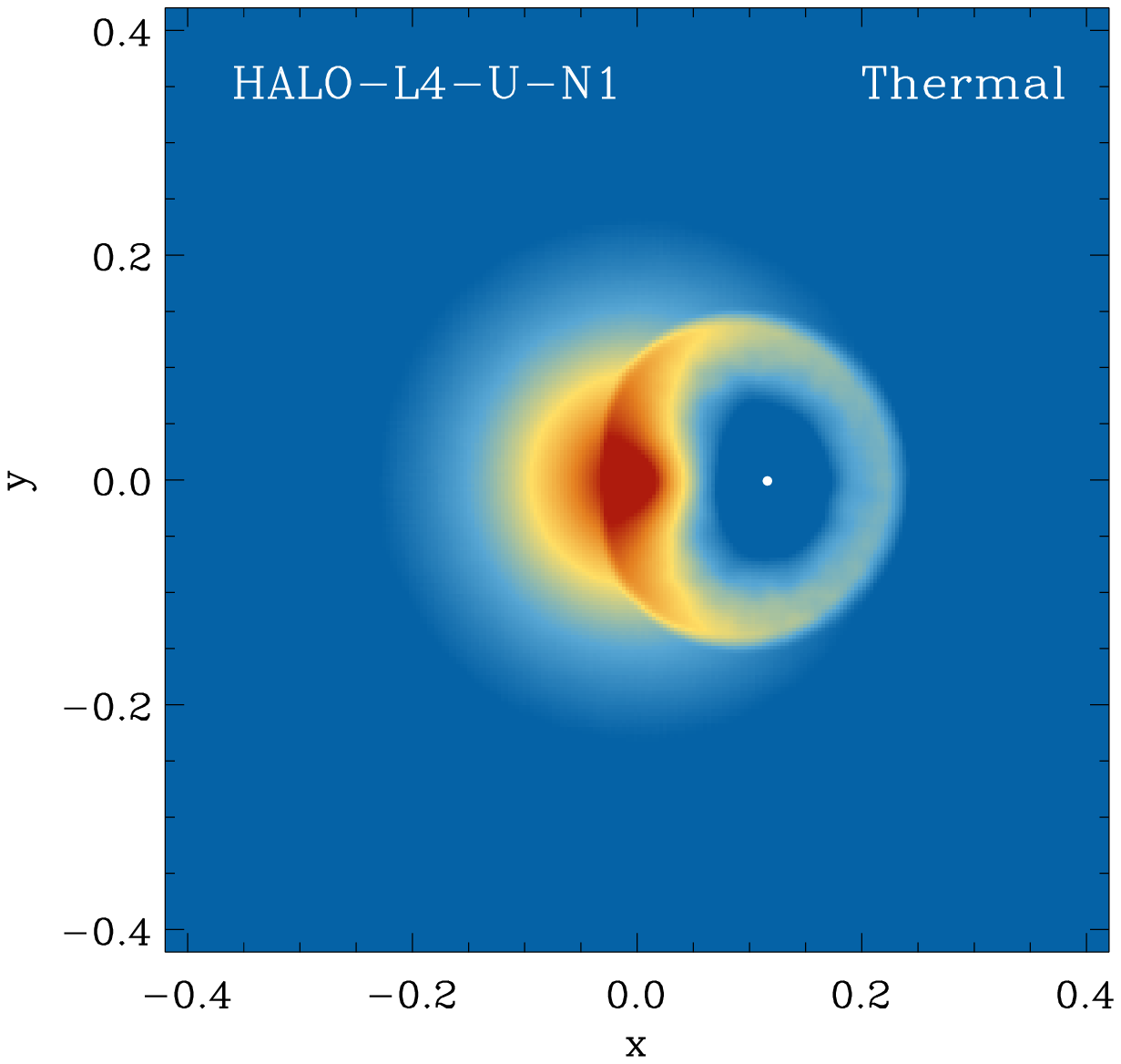}
    \includegraphics[trim = 8mm 0mm 4mm 8mm, clip, width=0.23\textwidth]{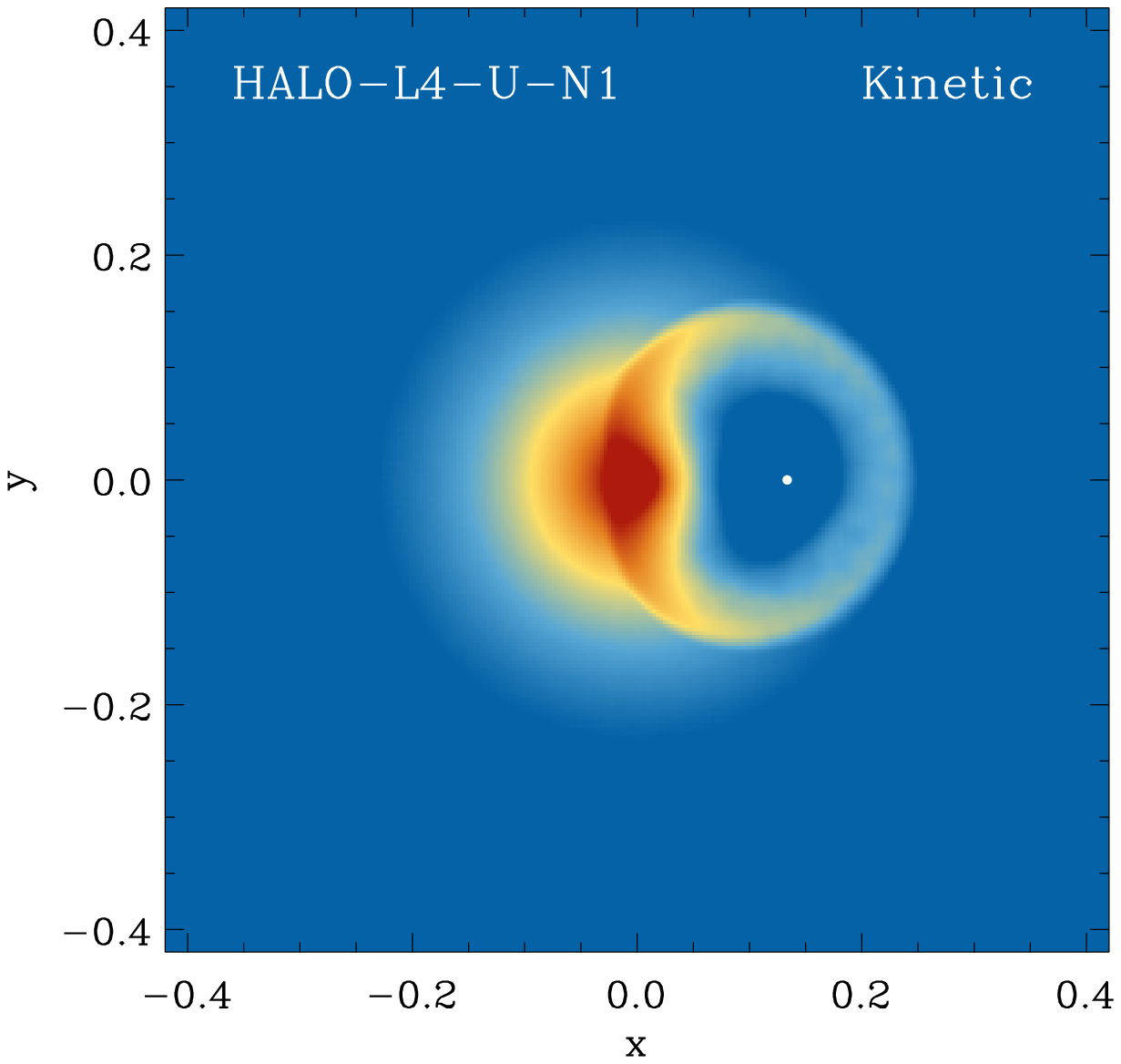}
  \end{center}
  \caption{The off-centre explosion in a self-gravitating gas sphere for thermal and kinetic energy injection when time-step limiter
      and update are applied. The plots are as in Fig.~\protect\ref{fig:evrwrong}. 
      \textit{Top row:} the energy is injected on 32 particles in thermal (left panel) and kinetic (right panel) form. 
      Impacted particles are made active and the time-step of surrounding particles has been corrected. 
      The results are qualitatively identical, showing concordance of the two feedback methods. 
      \textit{Bottom row:} all available energy is injected into one particle. 
      The results excellently agree with each other and with the cases in which the energy is injected into 32 particles.
      \textit{Time animation available as supporting online material.}
      \label{fig:evrcorrect}}
\end{figure}

The mass-weighted background internal energy at the position of the
explosion is $u=1.75$, whereas the mass-weighted velocity is
$v=4.43\times 10^{-2}$. The average internal energy and velocity are
measured in a spherical shell of thickness 0.004 and average radius 0.05.  In
the case of a single heated particle, the jump in energy would be
$\simeq 5\times 10^5$ times its initial value.  If only one particle
is kicked, its momentum change would be $\simeq 3\times 10^4$ times
the background initial value.

We compute the energy conservation as the relative variation of the total injected energy as in Sec.~\ref{sec:concordance}, and including the potential energy.
The simulation parameters together with the energy conservation estimate and the running time are listed in Table~\ref{table:evrardstat} 
where the reference simulations are highlighted in bold.

\subsubsection{Simulation results}
\label{subsec:evrardresults}

\smallskip

We show in Fig.~\ref{fig:evrwrong} the off-centre explosion in a self-gravitating gas sphere 
for thermal (left column) and kinetic (right column) energy injection.  
In all panels we show the projected gas density at time $t=0.02$ in a slice of thickness $\Delta z=0.1$ centred on the box centre. 
The white dots mark the position (at the same time) of all particles that initially received the energy. 
We first note the effect of the pressure and density gradients. 
In all pictures, the heated/kicked particles have moved from the location of energy injection along the direction of decreasing gradients (x-axis).

In the upper panels, we see the result of correcting the feedback scheme by making impacted particles active. 
Although heated/kicked particles are integrated accurately, their neighbours do not react soon enough. 
Indeed, before neighbours are active again, heated particles accelerate along the decreasing pressure gradients, 
and kicked particles build thermal energy through shocks. 
Once the neighbours feel the energy perturbation, they receive large accelerations. 
Integrating over a too long time-step gives them unrealistic momentum values, 
and in a few steps they are ejected from the vicinity of the explosion. 
Without the use of the limiter, the propagation of the effect from neighbour to neighbour
builds up energy to such levels that its conservation is violated by 
$\sim 57,000\:\%$ and $\sim 170,000\:\%$ for thermal and kinetic energy injection, respectively. 
It is interesting to see that both feedback implementations lead to the complete disruption of the halo atmosphere, which lost its initial spherical shape. 
This confirms the results of SM09 about the problems encounter with the standard individual time-stepping scheme and shows the importance of ensuring a smooth transport of the information from particle to particle.

In the lower panels of Fig.~\ref{fig:evrwrong}, we show the tests using the time-step limiter but without the time-step update. 
Though the limiter factor is set to the conservative value of two, the energy conservation is still largely violated.%

With thermal feedback (bottom-left panel), we recover the behaviour discussed in Sec.~\ref{subsec:individualsedovsetup}. 
The blast wave expands to large radii because thermal energy is overproduced within the hot bubble. 
Indeed, energy conservation is violated by $\sim3,000$\%, and the total thermal energy of the system is boosted to a factor of 20 or more the input one. 
At the final time, half of the thermal energy has been converted into momentum, and the total kinetic energy is two thirds of the total. 
Simulations of strong BH and SN feedback should be carefully checked against this behaviour. 

If kinetic energy is injected (bottom-right panel), the picture is similar to that discussed in Sec.~\ref{subsec:individualsedovsetup}. 
Inter-particle crossing is clearly recognisable in the picture. 
The particles with highest velocities (the closest to the injection position) have crossed several smoothing lengths and shock-heated the gas farther away. 
We note that within the blast wave several small bubbles are created, and dense gas is entrained, possibly decreasing the shell mass. 
Though less severe than the thermal case, energy conservation is violated by $\simeq20\%$, which, 
in cosmological simulations may lead to overestimating the effect of SN winds.
In summary, we showed that neither the update nor the limiter of impacted particles time-step are able by themselves 
to conserve energy within a reasonable accuracy level.

We show in Fig.~\ref{fig:evrcorrect} the off-centre explosion test when both the time-step limiter and update are applied (all plots are as in Fig.~\ref{fig:evrwrong}). 
We show in the upper panels the tests with $n_{\rm [h,k]}=32$, and in the lower ones the test with $n_{\rm [h,k]}=1$.
The agreement between the same model with different $n_{\rm [h,k]}$, and between different models, is striking. 
The evolution of the hot bubble is very similar in all cases, expanding and buoyantly moving from the place of energy injection to larger radii. 
The reference simulations with the global time-stepping integration show a conservation error $\lesssim2\%$, 
whereas the reference simulations with individual time-steps ($f_{\rm step}=4$, $n_{\rm [h,k]}=32$) have conservation error $\lesssim3\%$. 
We warn the reader that the gravitational force calculation is accurate at $\sim1\%$ level and could contribute to the errors listed in Table~\ref{table:evrardstat}.

Nonetheless, the lower panels of Fig.~\ref{fig:evrcorrect} show that concordance of thermal and kinetic feedback
is achieved even in the case of heating/kicking just one particle with
all the available energy. 

\begin{figure}
  \begin{center}
    \hspace{-4mm}\includegraphics[trim = 0mm 4mm 0mm 4mm, clip, width=0.46\textwidth]{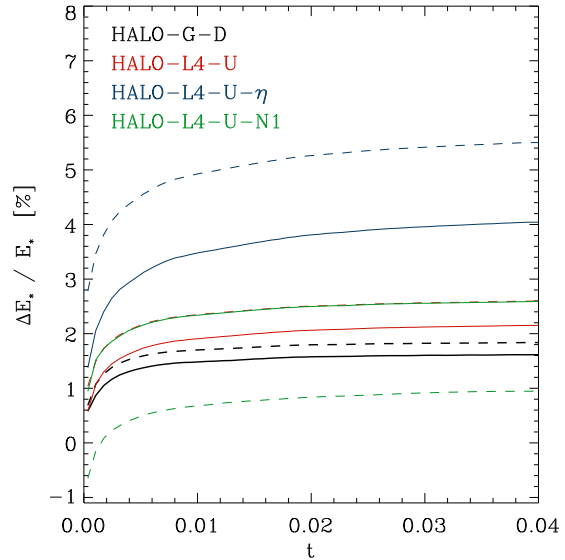}
  \end{center}
  \caption{Time evolution of the energy conservation for the most relevant tests of the explosion in a gas halo. 
    Solid lines show the results for the thermal injection of energy. 
    Dashed lines represent the evolution of the kinetic feedback runs.
    \label{fig:evrardenergy}}
\end{figure}

The time evolution of the energy conservation relative error is plotted in Fig.~\ref{fig:evrardenergy}. 
We plot some of the most relevant thermal (solid lines) and kinetic (dashed lines) feedback tests. 
Each line colour refers to simulations that were run with the same numerical parameters.
The agreement between kinetic and thermal feedback is excellent for the global time-stepping scheme (black lines), 
proving that a prompt response of the system to the energy perturbation is necessary to achieve concordance of the two methods.
Applying the proposed time-step scheme (red lines, reference model), the integration slightly looses accuracy, 
but simulations get a considerable speed up by a factor of $\sim8$ (see Table~\ref{table:evrardstat}). 
The agreement between the two feedback schemes is again excellent.

We show in blue the tests with $\eta=0.025$.
Given the lower time integration accuracy, energy violation
naturally reaches higher values, but the ratio of relative errors is similar to that of the reference tests. 
We see that the dashed blue curve goes to higher values right after the explosion. 
That shows that it is crucial to properly capture the conversion of momentum through
viscous forces at the very earliest time. At the final time, the relative error is 4 and $5.5\:\%$ for thermal and kinetic schemes, respectively.

Finally, simulations where only one particle is heated/kicked (green
lines) also show a good concordance of feedback methods. However,
injecting kinetic energy leads this time to a loss of total energy.
This can be explained because all particles close to the explosion are
inactive when energy is injected. Since all the energy is
concentrated, the kicked particle has the time to move for a few steps
while decelerated by viscous forces and before the surrounding medium
reacts. During this time, there is no energy transfer to neighbours which lead to the loss of energy.
This problem could be solved by making all neighbours active at the
same time of the impacted particles. We have tested this hypothesis,
and obtained again the concordance. In any case, energy is still
conserved with good accuracy.

\bigskip

\section{Conclusions}
\label{sec:discuss}

\bigskip

In this work we investigated the concordance of thermal and kinetic feedback methods in the absence of cooling. 
We numerically demonstrated that the two methods are equivalent using simulations of Sedov's explosion with a global time-step integration scheme. 
We argued that this result is expected from the hypothesis of Sedov's problem: a large amount of energy is instantaneously injected in a small volume. 
The blast physical properties are given by the integration of hydrodynamics conservation laws which also give the fraction of energy in thermal 
(72\%) and kinetic (28\%) form.

We focused on the conservation of input energy which we achieved within 1\% in the reference simulation. 
In test simulations with varying input energy, the energy conservation relative error flattens with time to roughly the same value, 
and is independent of the energy for a constant artificial viscosity parameter. 
Decreasing the artificial viscosity parameter results in a larger conservation error. 
We also investigated the evolution of the thermal and kinetic fractions of the input energy. 
We found excellent agreement with the expected values derived from Sedov's analytic solution.

Using the global time-step scheme in cosmological simulations is computationally expensive.
To reduce the computational costs most of the numerical codes use an adaptive, individual time-step scheme.
SM09 showed that to achieve good energy conservation of feedback energy a time-step limiter is necessary.
We implemented a similar limiter in \gadgettwo{} taking into account time-step synchronisation and hierarchy, and performed simulations of the Sedov's blast wave.
We showed that the limiter does not give accurate energy conservation if the explosion energy is injected at random time 
and a maximum value of the time-step for ambient gas particles is not enforced.

We proposed a solution to the problem: the system must not only propagate the information rapidly but should also promptly react to the energy input. 
That is done by updating the time-step of particles receiving energy accordingly to their new hydrodynamical state.

We tested the modified time integration scheme on the Sedov's blast wave, and obtained good energy conservation accuracy and concordance of methods. 
We applied both the limiter and the time-step update to a more realistic case of an explosion in a self-gravitating gas halo, 
and showed that concordance and accuracy are achieved also in the extreme case where all the available energy injected into one particle.

In the test simulations presented here, we did not consider radiative cooling, 
but we are aware of the potential breaking of concordance when such processes become important.
Cooling being a non-linear physical process, 
the amount of radiated energy before reaching energy partition will depend on the specific feedback scheme and the consequent gas temperature evolution. 
In the thermal feedback case, the bubble temperature decreases by adiabatic expansion and conversion of thermal energy into momentum.
In the kinetic case, the temperature increases through shocks. It is thus important to reach the same energy configuration before any significant cooling loss.
As mentioned by \cite{DallaVecchia2008}, if the gas is heated to a temperature where the cooling time becomes longer than the local dynamical time, 
radiative over-cooling may be prevented. These considerations will be addressed in more detail in an upcoming work.

\bigskip

We summarize this work as follows:

\smallskip

\begin{itemize}
\item concordance of thermal and feedback methods arises by accurate time integration and is expected from theoretical arguments;
\smallskip

\item keeping all other numerical parameters fixed, the maximum energy conservation relative error is roughly constant when varying the input energy, and at the same level with both feedback schemes;
\smallskip

\item high artificial viscosity coefficient enables a fast conversion of thermal energy to momentum, 
and hence permit to numerically converge to a stable solution in a shorter time;
\smallskip

\item concordance and energy conservation can be achieved with a time-step limiter, provided that the system promptly reacts to the input of energy;
\smallskip

\item in cosmological simulations with strong kinetic and/or thermal feedback from SNe and BHs, one should take into account accurate energy conservation before inferring any results dependent on feedback processes.
\end{itemize}

\section*{Acknowledgements}

We thank the anonymous referee for his useful comments that helped clarifying technical parts of this paper.
We are grateful to Volker Springel for making \gadgettwo{} publicly available. 
We acknowledge Takayuki Saitoh for the detailed description on his simulation setup. 
The authors are grateful to S.~Khochfar, E.~Neistein and J.~Johnson for helping improving the writing of the paper.
We also thanks Tom Theuns and Joop Schaye for enlightening comments and discussions. 
CDV is supported by the Marie Curie Reintegration Grant FP7-RG-256573. 
All simulations were performed on the SFC cluster of the TMoX group at the Max Planck Rechenzentrum Garching.

\begin{figure*}
  \begin{center}
    \hspace{.0cm}\includegraphics[trim = 0mm 4mm 0mm 0mm, clip, angle=-90,width=0.32\textwidth]{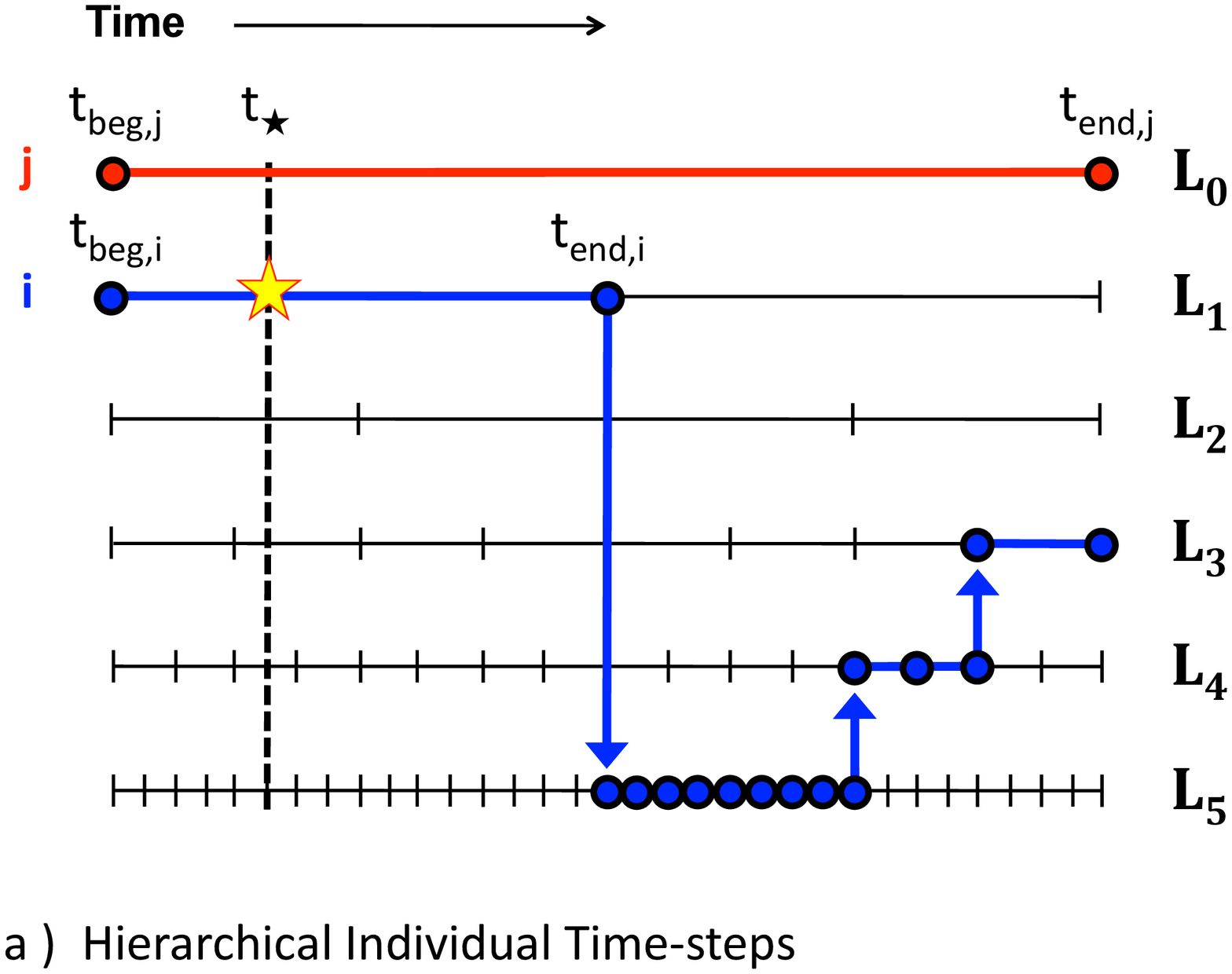}
    \hspace{.2cm}\includegraphics[trim = 0mm 4mm 0mm 0mm, clip, angle=-90,width=0.32\textwidth]{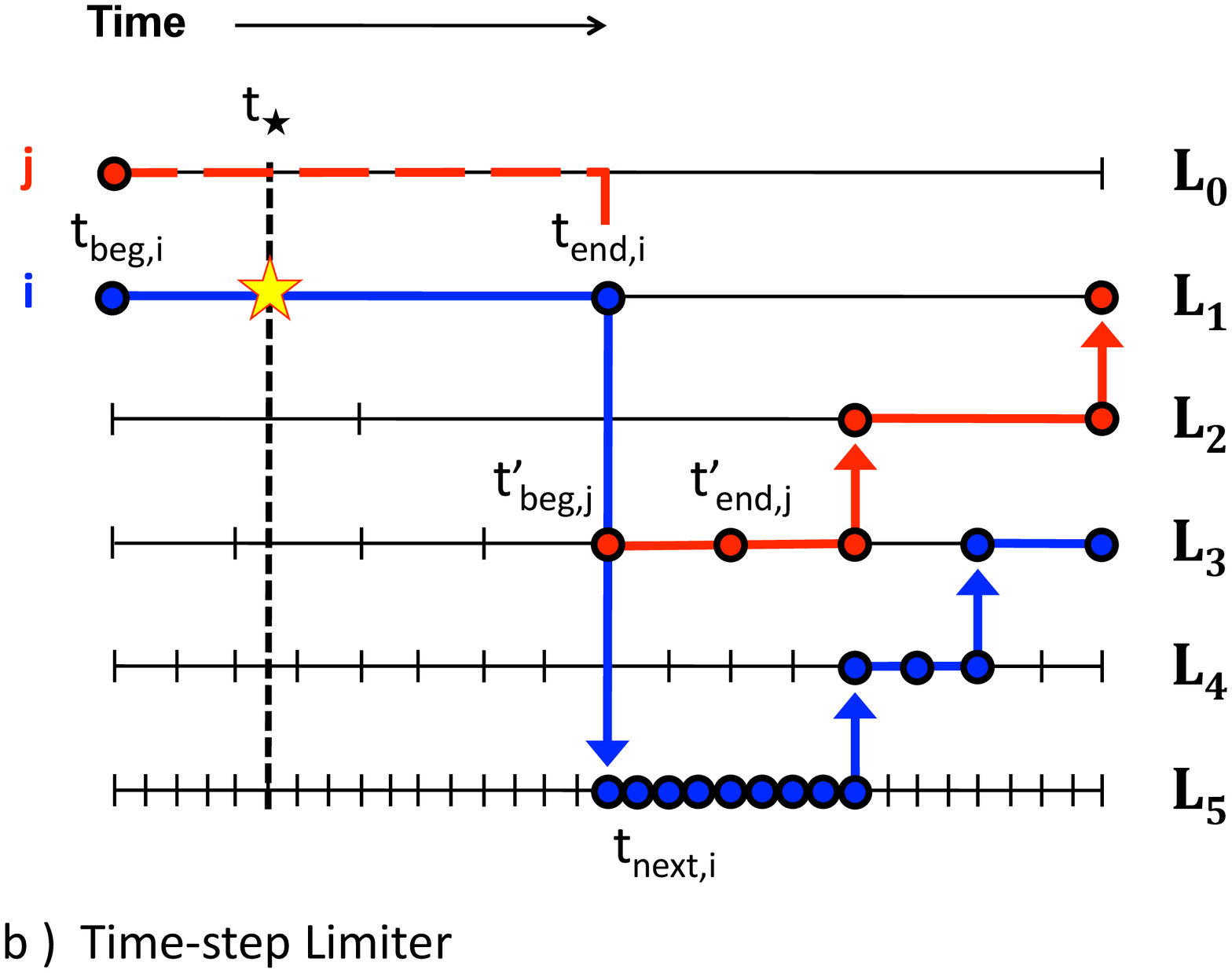}
    \hspace{.2cm}\includegraphics[trim = 0mm 4mm 0mm 0mm, clip, angle=-90,width=0.32\textwidth]{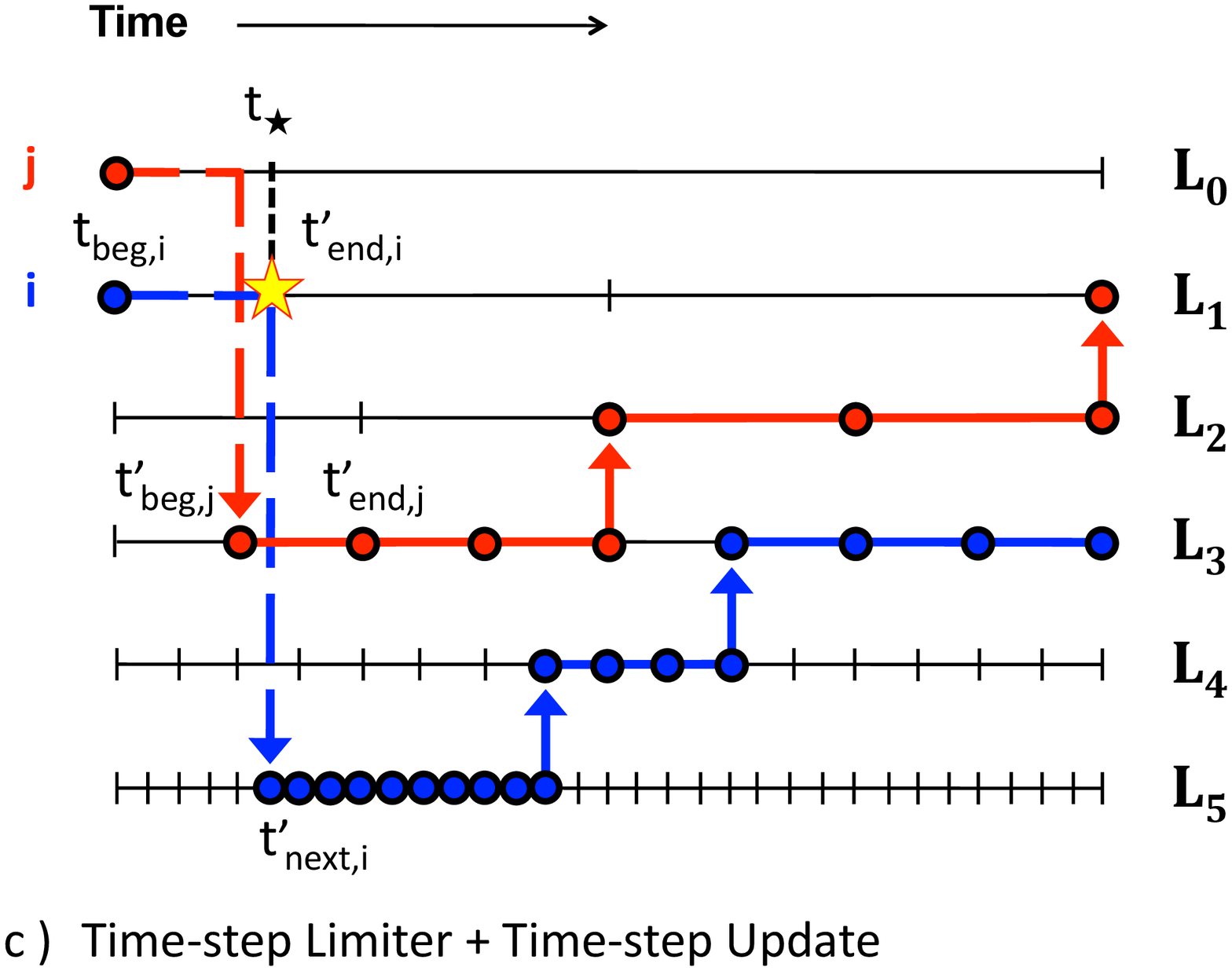}
  \end{center}
  \caption{Timelines for a synchronous time integration scheme. Energy is given to particle $i$, and particle $j$ is one of its neighbours. Solid lines represent time intervals where the state of particles is re-computed, while long-dashed lines correspond to ``fictitious'' steps. Circles represents the ``active'' times of particles. Labels given on the right side of each sketch show the levels of the time-bin hierarchy. a)~Individual time-steps: particles are only able to adapt their time-step when the simulation time equals their end-step time (when they become active). Therefore, heated/kicked particles may complete a significant number of steps before their neighbours become active as well. b)~Time-step limiter: neighbouring particles communicate to each other the length of their time-steps, and keep the ratio of long to short time-steps not larger than a factor of four ($f_{\rm step}=4$). However, particles still need to be active before their neighbours adjust their time-step accordingly to the limiter criterion. c)~Time-step limiter and time-step update: in the scheme introduced in this work, we make sure that heated/kicked particles become active at the time of energy injection, and adjust their time-step accordingly to the amount of energy they receive. When applied in combination with time-step limiter, impacted particles and their neighbours can promptly react to the change of energy in the medium.\label{fig:timelines}}
\end{figure*}

\begin{appendix}
\bigskip

In the following appendices we describe all the relevent part of the time integration implementation of \gadgettwo{}. 
The reader who is not familiar with the leapfrog integration may read Appendix~\ref{app:leapfrog}. 
There, we also remind about the individual time-step hierarchy on which the leapfrog is applied.
In Appendix~\ref{app:dtcriteria} we review the two criteria used to define the individual time-steps 
and explain the modification we made to match our need of accuracy.
In Appendix~\ref{app:timelimiter}, we proceed to the description of the implementation of the time-step limiter in \gadgettwo{}.
Finally, we discuss in Appendix~\ref{app:timeupdate} our time-step update which enables to achieve a prompt response of the system to any type of energy perturbation.

\vspace{-.4cm}
\section{Leapfrog integrator}
\label{app:leapfrog}

For a given time interval $\Delta t=t_{n+1} - t_{n}$, the KDK leapfrog integrator used in \gadgettwo{} \citep[first presented in this form by][]{Quinn1997} 
is defined by the following time evolution operator:
\begin{equation}
\tilde{U}\bigl(\Delta t\bigr) = K'\biggl(\frac{\Delta t}{2}\biggr)\,
D\bigl(\Delta t\bigr)\, K''\biggl(\frac{\Delta t}{2}\biggr)\,,
\label{eq:evolop}
\end{equation}
where the \textit{kick} ($K$) and \textit{drift} ($D$) operators are used to evolve both the dynamical and hydrodynamical quantities that characterise any particle as follows:
\begin{align}
  K'\left(\frac{\Delta t}{2}\right): &
  \label{eq:kick1}
  \begin{cases}
    \mathbf{x}_{n} &\rightarrow\quad \mathbf{x}_{n} \\
    \mathbf{v}_{n} &\rightarrow\quad \mathbf{v}_{n+\frac{1}{2}} = \mathbf{v}_{n} + \mathbf{a}_{n} \frac{\Delta t}{2} \\
    \mathrm{P}_{n} &\rightarrow\quad \mathrm{P}_{n} \\
    \mathrm{S}_{n} &\rightarrow\quad \mathrm{S}_{n+\frac{1}{2}} = \mathrm{S}_{n} + \dot{\rm S}_{n} \frac{\Delta t}{2}
  \end{cases}
  \\ \nonumber \\
  D(\Delta t): &
  \label{eq:drift}
  \begin{cases} 
    \mathbf{x}_{n}  &\rightarrow\quad \mathbf{x}_{n+1} = \mathbf{x}_{n} + \mathbf{v}_{n+\frac{1}{2}} \Delta t \\
    \mathbf{v}_{n+\frac{1}{2}} &\rightarrow\quad \mathbf{v}_{n+\frac{1}{2}} \\
    \mathrm{P}_{n}  &\rightarrow\quad \mathrm{P}_{n+1} = \left( \mathrm{S}_{n+\frac{1}{2}} + \dot{\rm S}_{n} \frac{\Delta t}{2} \right)\, \rho_{n}^{\gamma} \\
    \mathrm{S}_{n+\frac{1}{2}} &\rightarrow\quad \mathrm{S}_{n+\frac{1}{2}} 
  \end{cases}
  \\ \nonumber \\
  K''\left(\frac{\Delta t}{2}\right): &
  \label{eq:kick2}
  \begin{cases}
    \mathbf{x}_{n+1} &\rightarrow\quad \mathbf{x}_{n+1} \\ 
    \mathbf{v}_{n+\frac{1}{2}} &\rightarrow\quad \mathbf{v}_{n+1} = \mathbf{v}_{n+\frac{1}{2}} + \mathbf{a}_{n+1} \frac{\Delta t}{2} \\
    \mathrm{P}_{n+1} &\rightarrow\quad \mathrm{P}_{n+1} \\
    \mathrm{S}_{n+\frac{1}{2}} &\rightarrow\quad \mathrm{S}_{n+1} = \mathrm{S}_{n+\frac{1}{2}} + \dot{\rm S}_{n+1} \frac{\Delta t}{2} 
  \end{cases}
\end{align}
In the above equations, $\mathbf{x}_n$, $\mathbf{v}_n$, $\mathbf{a}_n$, $\mathrm{P}_{n}$, $\rho_{n}$, $\mathrm{S}_n$ and $\dot{\rm S}_n$ are respectively, 
position, velocity, acceleration, pressure, density, entropic function and its rate of variation at time $t_n$. 
It is interesting to note here that the acceleration and the rate of entropy change need to be computed before the $K$ operator is applied 
(which implies an update of the system), while the drifting is a free-motion operation that uses the velocity and entropy values from the previous kick.

In cosmological simulations of galaxy formation, where a large range of time scales is required, 
the use of an individual time-step scheme enable to reduce the number of force computations per step 
to only those needed by a fraction of the system. 
The KDK integrator is then used over small time-steps, $\delta t$, for those particles, 
while the other particles experience fewer kicks over longer steps. 
However, given the pair-wise nature of the forces, the potential is required to
evolve on the shorter time scale, in order to preserve the accuracy of
the force computation. As long as the larger time-steps can be
subdivided into several shorter ones, $\Delta t = \sum_k \delta t_k$,
the splitting of the drift operator for the long time-step particles
ensure that the positions of all particles are updated for any force
calculation:
\begin{equation}
  \label{eq:DriftSum}
  D\left(\Delta t\right) = 
  \sum_k D\left(\delta t_k\right)\,.
\end{equation}

If the individual time-step scheme allows to integrate each particle
according to its local dynamical state, one side effect is that the
positions of all the particles of the system need to be drifted in
order to compute the forces at a given time. To optimise the
efficiency of the leapfrog integrator, one has to impose that several
particles share exactly the same simulation time. This can be
achieved by discretising the time-steps, as proposed by
\cite{Hernquist1989} and \cite{Makino1991}, in a hierarchy of power of
two subdivisions of the global simulation time, $t_{\rm glob}$:
\begin{equation}
\delta t_k = \frac{t_{\rm glob}}{2^{k}}\,,
\end{equation}
where $k\ge0$ is an integer that define the level of the corresponding
time-bin. Particles are allowed to move to a smaller time-step, but
can only jump up to the next larger time-step every other step (see
panel \textit{a} of Fig.~\ref{fig:timelines} for an illustration). An
additional advantage of the hierarchical scheme is that the previous
constraint leads to the synchronisation between successive time levels
which helps to balance the distribution of particles into time-bins.

In \gadgettwo{}, the length of the particle time-step is given by two
variables storing the beginning, $t_{\rm beg}$, and ending, $t_{\rm
  end}$, time of the current particle step. All particles with
$t_{\rm end}$ equals the current simulation time are considered
\textit{active}.  This means that their dynamical and hydrodynamical
states are updated by the following operator:
\begin{align}
  A(t_{\rm end}): &
  \begin{cases}
    \mathbf{a}_{\rm beg}  & \rightarrow\quad \mathbf{a}_{\rm end} = \mathbf{a}^{\rm grav}_{\rm end} + \mathbf{a}^{\rm hydro}_{\rm end} \\
    \dot{\rm S}_{\rm beg} & \rightarrow\quad \dot{\rm S}_{\rm end} 
  \end{cases}
\end{align}
where the acceleration, $\mathbf{a}$, describes the motion of
the fluid element, and takes also into account the artificial viscous
forces generated by the shocks that develops into the medium. As a
consequence of the local viscosity, the momentum of the gas is
converted into entropy at a rate given by $\dot{\rm S}$. However,
before applying the operator $A$, some other quantities need to
be updated, such as the smoothing length, and the local density and
pressure. The reader should refer to the original \gadgettwo{} paper
for more details about the specific expression of these quantities.

To conclude this reminder of the time integration scheme used in
\gadgettwo{}, we want to remark that the two kick operators $K'$ and
$K''$ given in expression \ref{eq:kick1} and \ref{eq:kick2}, will use
the acceleration and the entropy rate computed at the same time, 
$t_{\rm end}$, when applied in two successive steps. The
implementation of the leapfrog method in \gadgettwo{} makes use of this
property to save some computational operations. At the end of each
active step, the length of the next time-step, $\Delta t_{\rm next}$, is computed before
applying the kick operator from the current half-step to the next
half-step. Therefore, the time evolution operator can be rewritten
as:
\begin{eqnarray}
  \tilde{U} & = & D(t_{\rm beg}\longrightarrow t_{{\rm end}}) \\ \nonumber
  && A(t_{\rm end})\,
  K\left( t_{\rm beg}+ \frac{\Delta t_{\rm now}}{2} \longrightarrow t_{\rm end}+\frac{\Delta t_{\rm next}}{2} \right)\,,
\label{eq:evolop2}
\end{eqnarray}
where $D$ is applied at any simulation step that sub-divide $\Delta
t_{\rm now}$, while both $A$ and $K$ are only evaluated at the end of
the individual steps for active particles.

\section{Time-step criteria}
\label{app:dtcriteria}

Time-steps are chosen as the smallest given by the Courant criterion and the acceleration criterion.
The Courant time-step criterion is defined by the signal velocity method, $v_{{\rm sig},i}$, \citep{Monaghan1997}
which estimates the speed at which the information propagates at the spatial resolution scale,
\begin{equation}
\Delta t_{{\rm C},i}=C\frac{2h_i}{v_{{\rm sig},i}}.
\label{eq:courdt}
\end{equation}
where $C$ is the Courant factor and $h_i$ the smoothing length.\footnote{Note that in 
\gadgettwo{} the smoothing length $h$ is the size of the SPH kernel, whereas other authors define the kernel of size $2h$. 
Hence, the fiducial value of the Courant factor used in \gadget is half of the one found in other SPH codes.}
\smallskip

The acceleration time-step criterion is defined in \gadgettwo{} as
\begin{equation}
\Delta t_{{\rm acc},i}=\sqrt{\eta\, \frac{2\,\epsilon}{a_i}}\,,
\label{eq:accdt}
\end{equation}
where $\eta$ is the accuracy parameter, $\epsilon$ is the gravitational softening, 
and $a_i$ is the particle acceleration.\footnote{In pure hydrodynamic simulations, the acceleration is given only by hydrodynamical forces while, 
in simulations that also include self-gravity, the particle acceleration is the sum of the contributions from gravitational and hydrodynamical forces.}
Though this criterion gives accurate integration in most applications, it is crucial that the initial, 
large hydrodynamical accelerations, encountered under strong energy perturbations, are properly captured.
However, since we allow the individual smoothing length $h_i$ to be smaller than the gravitational softening, 
the above criterion could overestimate the length of the time-step. 
Following \cite{Monaghan1997}, we thus modified the acceleration criterion for SPH particles as follows:
\begin{equation}
\Delta t_{{\rm acc},i}=\sqrt{\eta\, \frac{2\,\mbox{min}(h_i,\epsilon)}{a_i}}\,.
\label{eq:accdtnew}
\end{equation}
\smallskip

In addition, as reported by \cite{Wuyts2010}, the value of the accuracy coefficient, $\eta$, 
may play a role in capturing strong shocks, and hence in the convergence of the results. 
This motivated the accuracy tests presented in Sec.~\ref{sec:sedov} and \ref{sec:evrard} 
which made us choose $\eta=0.0025$ as fiducial parameter for this work.

\section{Time-step limiter}
\label{app:timelimiter}

We generalise the implementation of the time-step limiter in order 
to deal with energy injection happening at random time and affecting active and/or inactive particles.
We describe here the implementation of the time-step limiter into
\gadgettwo. The main idea behind the limiter is to ensure that every
neighbouring particle $j$ is integrated over a time-step that is
shorter than or equal to a multiple of particle $i$ time-step:
\begin{equation}
\label{eq:limiter}
\Delta t_j \le f_{\rm step} \Delta t_i\,.
\end{equation}

In the context of the kick-drift-kick (hereafter, KDK) leapfrog integration presented in Appendix~\ref{app:leapfrog}, 
particle $i$, which experiences a change of its energy content in the middle of its current time-step, 
will only be able to react to the explosion when it becomes active. This is illustrated in Fig.~\ref{fig:timelines}~a).
At this time, and assuming that the energy change is carefully taken into account (see discussion in the next Appendix), 
the particle \textit{next} time-step, $\Delta t_i$, will be brought to a finer level allowing a more accurate computation of the hydrodynamics.
Later on, when the medium permits it, the particle time-step will eventually increase according to the synchronisation scheme. 
As shown in Fig.~\ref{fig:timelines}~a), the neighbouring particle $j$ will have to wait a long time before becoming active and acknowledging the energy change. 
Therefore, to make sure that all the neighbouring particles will react to the explosion, particle $i$ has to communicate them its next time-step.

The implementation of the limiter is illustrated in Fig.~\ref{fig:timelines}~b) and proceeds as follows. 
We consider particle $i$ which receives energy at time $t_{\star}$, and particle $j$ which is one of its neighbours. 
When active, particle $i$ informs neighbour $j$ of its next time-step. 
If the neighbouring particle $j$ is active and the inequality in Eq.~\ref{eq:limiter} is not satisfied,
its time-step will be shortened to the value $\Delta t_j'=f_{\rm step} \Delta t_i$. 
If particle $j$ is inactive, the time-step limiter criterion will ensure that it will be recomputed no later than:
\begin{equation}
t_{\rm lim} = t_{{\rm end},i} + f_{\rm step}\Delta t_i\,,
\label{eq:limitercond}
\end{equation}
where $t_{{\rm end},i}$ is the end-time of particle $i$ and $\Delta
t_i$ is its next time-step.\footnote{The time $t_{{\rm end},i}$ is
  actually the current simulation time, since particle $i$ is active.}
If particle $j$ time-step ends later than the limiter time, $t_{{\rm
    end},j} > t_{\rm lim}$, we change it to the \textit{new} value:
\begin{equation}
t_{{\rm end},j}' = t_{{\rm beg},j} + k f_{\rm step} \Delta t_i\,,
\end{equation}
where $k$ is the largest integer that verifies the condition
\begin{equation}
t_{{\rm beg},j} + k f_{\rm step} \Delta t_i \le t_{\rm lim}\,.
\end{equation}
This ensures that particle $j$ is computed within $t_{\rm lim}$.  
A new time-step, $\Delta t_j'=f_{\rm step} \Delta t_i$, is thus assigned to particle $j$ and the beginning of the step becomes 
$t_{{\rm beg},j}' = t_{{\rm end},j}' - \Delta t_j'$. 
This is done to preserve the synchronisation in the time-step hierarchy. 
Finally, because both the beginning and end-step of particle $j$ have been modified, 
velocity and entropy must be made consistent with the new time-step, 
by interpolating their values from the middle of the original time-step to the middle of the new one. 
This can be done by applying the following kick operator (see Appendix~\ref{app:leapfrog})
\begin{equation}
  K\left(t_{{\rm beg},j} + \frac{\Delta t_j}{2} \longrightarrow
  t_{{\rm beg},j}'+\frac{\Delta t_j'}{2} \right)\,.
\end{equation}
Since the particle is inactive, the acceleration and the rate of entropy change used in the interpolation are the ones computed at the beginning of the previous active step.

Regardless of the implementation of the limiter, we can clearly see in Fig.~\ref{fig:timelines}~b) that, if the explosion occurs in the middle of the particle $i$ time-step, 
a considerable amount of time passes before both particles are recomputed accordingly to this event. 
In order to prevent this undesirable behaviour, we present in the next section our suggestion to correct the timeline of impacted particles.

\section{Time-step update}
\label{app:timeupdate}

The basic idea behind our time-step update can be divided into two
specific actions: i)~to make sure that the computation of the step
following the explosion takes into account the local change of
the energetics; ii)~to ensure that impacted particles react as soon
as the energy is injected.

In the following, we consider an instantaneous injection of energy,
where the first time-step following the explosion event defines how
far the information can travel before the medium starts to absorb the
energy. It is thus crucial that the time-step of impacted particles
does not violate either the Courant or the acceleration criterion.  In
our implementation, we therefore update the maximum signal velocity
and compute the acceleration of impacted particles just after the call
of the feedback routine. The acceleration will be used only in the
calculation of the time-step, and does not substitute the one computed
at the beginning of the step. Given the pair-wise nature of the
hydrodynamical forces, the time-steps of active particles in the
neighbourhood of the energy injection area are conveniently updated at
the same time. The computational cost of an additional neighbour
search is only a negligible fraction of that required for updating the densities
and the hydrodynamical forces. Finally, this method has the advantage
of naturally taking into account the effect of multiple sources of
feedback.

The second action that is needed for a prompt response of the medium,
is to make active all the particles inside the explosion area.  As
shown in Fig.~\ref{fig:timelines}~c), this rather trivial operation
allows particle $i$ to shrink its time-step straight after the
explosion time, $t_{\star}$. We can see from the figure that, when
used jointly with the time-step limiter, both the impacted particles
and their neighbours are now integrated on time scales that are able
to correctly handle the upcoming evolution of the blast wave.

To briefly describe the implementation, we refer again to the KDK
leapfrog scheme in
Appendix~\ref{app:leapfrog}. At the time of explosion $t_{\star}$, the
positions of all particles have already been drifted with the D
operator. Since we want the particles to be recomputed just after the
explosion, we have to apply again the kick operator to complete a
``fictitious'' step. The main difference with the kick correction
used for the limiter being that the operator is here applied back in
time:
\begin{equation}
  K\left( \frac{t_{{\rm beg},i} + t_{{\rm end},i}'}{2} \longleftarrow \frac{t_{{\rm beg},i} + t_{{\rm end},i}}{2} \right)\,,
\end{equation}
where $t_{{\rm end},i}$ is the previous end-step time of particle $i$ and
$t_{{\rm end},i}'=t_{\star}$ is the current simulation time.

Making the particle active at the explosion time allows particle $i$ to recompute its next time-step $\Delta t_{{\rm next},i}'$ according to the Courant criterion. 
Since this particle has not really been updated, the kick which closes this fictitious step, ${\rm K''}$, 
will use the acceleration and rate of entropy change computed at time $t_{{\rm beg},i}$. 
This is consistent with an instantaneous injection of energy, and prevents any loss of it. 
As illustrated in Fig.~\ref{fig:timelines}~c), the heated/kicked particle is now able 
to handle its new energy content more accurately and can inform its neighbours immediately through the time-step limiter criterion.

\end{appendix}

\end{document}